\documentclass[journal,twoside]{IEEEtran} 

\usepackage{cite}
\usepackage{amsmath,amssymb,amsfonts}
\usepackage{array}
\usepackage[noend]{algpseudocode}
\usepackage[noend,ruled,vlined,linesnumbered]{algorithm2e}
\usepackage{graphicx}
\usepackage{textcomp}
\usepackage{xcolor}
\usepackage{flushend}
\usepackage{multirow}
\usepackage{booktabs}
\usepackage{tabularx}
\usepackage{ragged2e}
\usepackage[final]{microtype}
\usepackage{comment}
\usepackage[alpha]{mdpn} 
\ifCLASSOPTIONcompsoc
  \usepackage[caption=false,font=normalsize,labelfont=sf,textfont=sf]{subfig}
\else
  \usepackage[caption=false,font=footnotesize]{subfig}
\fi

\newcolumntype{L}{>{\RaggedRight\arraybackslash}X} 

\def\BibTeX{{\rm B\kern-.05em{\sc i\kern-.025em b}\kern-.08em
    T\kern-.1667em\lower.7ex\hbox{E}\kern-.125emX}}
\begin{document}

\title{On the Robustness of Controlled Deep Reinforcement Learning for Slice Placement}

\author{\IEEEauthorblockN{Jose Jurandir Alves Esteves \IEEEauthorrefmark{1}\IEEEauthorrefmark{2}, Amina Boubendir \IEEEauthorrefmark{1}, Fabrice Guillemin \IEEEauthorrefmark{1} and Pierre Sens \IEEEauthorrefmark{2}}

\IEEEauthorblockA{\IEEEauthorrefmark{1}Orange Labs, France} \IEEEauthorrefmark{2}Sorbonne Universit\'e / CNRS / Inria, LIP6, France\\ \{josejurandir.alvesesteves, amina.boubendir, fabrice.guillemin\}@orange.com, pierre.sens@lip6.fr

\thanks{J.J. Alves Esteves is with Orange Labs, 92320 Chatilon, France and also with LIP6 -- Inria, Sorbonne Univ., 75005 Paris, France (e-mail: josejurandir.alvesesteves@orange.com).}
\thanks{A. Boubendir and F. Guillemin are with Orange Labs, 92320 Chatillon, France (e-mail: firstname.name@orange.com).}
\thanks{P. Sens is with LIP6 -- Inria, Sorbonne Univ., CNRS, 75005 Paris, France (e-mail: pierre.sens@lip6.fr).}
}

\markboth{IEEE Journal on Selected Areas in Communication}%
{ALVES ESTEVES \MakeLowercase{\textit{et al.}}: 
On the Robustness of Controlled Deep Reinforcement Learning for Slice Placement}

\maketitle

\begin{abstract}
 The evaluation of the impact of using Machine Learning in the management of softwarized networks is considered in multiple research works. Beyond that, we propose to evaluate the robustness of online learning for optimal network slice placement. A major assumption to this study is to consider that slice request arrivals are non-stationary. In this context, we simulate  unpredictable network load variations and compare two Deep Reinforcement Learning (DRL) algorithms: a pure DRL-based algorithm and a heuristically controlled DRL as a hybrid DRL-heuristic algorithm, to assess the impact of these unpredictable changes of  traffic load on the algorithms performance. We conduct extensive simulations of a large-scale operator infrastructure. The evaluation results show that the proposed hybrid DRL-heuristic approach is more robust and reliable in case of unpredictable network load changes than pure DRL as it reduces the performance degradation. These results are follow-ups for a series of recent research we have performed showing that the proposed hybrid DRL-heuristic approach is efficient and more adapted to real network scenarios than pure DRL.
\end{abstract}


\section{Introduction}

\IEEEPARstart{T}{he} promise of {network} Slicing is to enable a high level of customization of network services in future networks (5G and beyond) leveraged by  virtualization and software defined networking techniques. These key enablers  transform telecommunications networks into programmable platforms capable of offering virtual networks enriched by Virtual Network Functions (VNFs) and IT resources tailored to the specific needs of certain customers (e.g., companies) or vertical markets (automotive, e-health, etc.)\cite{3GPP,etsi}. From an optimization theory perspective, the Network Slice Placement problem can be viewed as a specific case  of Virtual Network Embedding (VNE) or VNF Forwarding Graph Embedding (VNF-FGE) problems \cite{survey_vnf_ra_2016,survey_vfnp}. It is then generally possible to formulate Integer Linear Programming (ILP) problems \cite{netsoft_2020}, which however turn out to be $\mathcal{NP}$-hard \cite{vne_np_hardness} with very long convergence time. 

With regard to network management, there are specific characteristics related to network slicing: slices are expected to share resources and coexist in a large and distributed infrastructure. Moreover, slices have a wide range of requirements in terms of resources, quality objectives and lifetime. 

In practice, these characteristics bring additional complexity as the placement algorithms need to be highly scalable with low response time even under varying network conditions. 

As an alternative to optimization techniques and the development of heuristic methods, Deep Reinforcement Learning (DRL) has recently been used in the context of VNE and Network Slice Placement \cite{p1,p2, p5, p3, p4, p8}. DRL techniques are considered as very promising since they allow, at least theoretically, the determination of optimal decision policies only based on experience \cite{sutton2018reinforcement}. However, from a practical point of view, especially in the context of non-stationary environments, ensuring that a DRL agent converges to an optimal policy is still challenging. 

As a matter of fact, when the environment is continually  changing, the algorithm has trouble in using the  acquired knowledge to find optimal solutions. The usage of the DRL algorithm online  can then become impractical. In fact, most of  existing works based on DRL to solve the Network Slice Placement or VNE problem assume a stationary environment, i.e., with constant network load. However, traffic conditions in real networks are basically non stationary with daily and weekly variations and subject to drastic changes (e.g., traffic storm due to an unpredictable event).   

To cope with traffic changes, this paper proposes a hybrid DRL-heuristic strategy called Heuristically Assisted DRL (HA-DRL)\cite{HA_DRL_TNSM}. We applied in \cite{cnsm_2021} this strategy in  an online learning scenario with periodic network load variations to show how this strategy can be used to accelerate and stabilize the convergence of DRL techniques in this type of non-stationary environment. As a follow-up of theses two studies, we focus in the present paper on a different non stationary scenario with stair-stepped  network load changes. The goal of the paper is to evaluate and show the robustness of the proposed strategy method in the case  of sudden and stair-stepped traffic changes.

The  contributions of the present paper are threefold: 
\begin{enumerate}
    \item We propose a network load model to describe network slice demand and adapt it to unpredictable network load changes;
    \item We propose a framework combining Advantage Actor Critic and a Graph Convolutional Network (GCN) for conceiving DRL-based algorithms adapted to the non-stationary case;
    \item We show how the use of a heuristic function can control the DRL learning and improve its robustness to unpredictable network load changes.
\end{enumerate}

The organization of this paper is as follows: In Section~\ref{sec:sota}, we review the related work. In Section~\ref{sec:network_model}, we describe the Network Slice Placement problem modeling. The learning framework for slice placement optimization is described in Section~\ref{sec:drl_proposal}. The adaptation of the pure DRL approaches and its control by using heuristic is introduced in Section~\ref{sec:aidedDRL}. The experiments and evaluation results are presented in Section~\ref{sec:evaluation}, while conclusions and perspectives are presented in Section~\ref{sec:conclusion}.

\section{Related Work Analysis \label{sec:sota}}

We provide in Section \ref{sec:drl_based} a summarized review of the existing DRL-based approaches for network slice placement. The interested reader may refer to \cite{HA_DRL_TNSM,cnsm_2021} for a more detailed and comprehensive discussion. In Section \ref{sec:robustness} we discuss recent works on robust slice placement algorithms.

\subsection{On DRL-based Approaches for Slice Placement \label{sec:drl_based}}

DRL has been recently applied to solving network slice placement and VNE problems. We divide these works into two categories on the basis of their algorithmic aspects: 1) pure DRL approaches \cite{p1,p2, p5, p3, p4, p8}, in which only the knowledge acquired by the learning agent via training is used as a basis for taking placement decisions; and 2) hybrid DRL-heuristic approaches \cite{HA_DRL_TNSM,quang2019deep,rkhami2021learn}, in which the placement decision computation is assisted by a heuristic method. 

The use of heuristics aims at increasing the reliability of DRL algorithms. However, most of these works are based on the assumption that the network load is static, i.e., slice arrivals occurs at a constant rate. To the best of our knowledge, the work we proposed in \cite{cnsm_2021} is the first attempt to evaluate an online DRL-based approach in a non-stationary network load scenario whereas  \cite{new_1} only considers offline learning.

In addition, in both \cite{cnsm_2021} and \cite{new_1} is assumed that network load has periodic fluctuations. In the present paper we study the behavior of the  algorithms proposed in \cite{cnsm_2021} in case of an unpredictable network load disruption.  

\subsection{On Robustness of Slice Placement Approaches \label{sec:robustness}}

The term robustness has different meanings depending on the field of application. In Robust Optimization (RO), robustness is related to the decision/solution itself. It is the capability of the algorithm solution of coping with the worst case  without losing feasibility \cite{bertsimas2006robust}. 

In Machine Learning (ML), specially in Deep Learning (DL), robustness is related to the learned model. It is the property of the model (i.e., Deep Neural Network (DNN)) that determines its integrity under varying operating conditions \cite{shafique2020robust}. 

The authors of \cite{al2021robustness} are the first to discuss robustness in the DRL context. They propose to use Genetic Algorithm to improve the robustness of a self-driving car application. Robustness is considered as the capacity of sustaining a high accuracy on image classification even when perceived images change and it is measured by Neuron Coverage (NC), i.e.,the ratio of the activated neurons in the DNN.  

There are only a few recent works on the robustness of slice placement procedures, most of them on RO \cite{marotta2017fast,marotta2017energy,reddy2016robust,baumgartner2017}. These works answer a  question different from the one we are investigating as they evaluate the robustness of the decision whereas we want to evaluate the robustness of the learning process. Despite their originality, the above approaches present some drawbacks, such as the lack of scalability of ILP, the sub-optimality of heuristic solutions, the fact that they consider offline optimization in which all slices to be placed are known in advance, and the fact that they are single objective optimization approaches, mainly focusing on energy consumption minimization. In this work, we propose to rely on a DRL-based approach in order to overcome ILP and heuristic drawbacks and consider multiple-optimization objectives.

To the best of our knowledge, paper~\cite{robust_1} is the only one to have proposed a DRL-based approach for slice placement and evaluated the learning robustness. However, the authors focus on evaluating the robustness of the DRL approach against random topology changes (e.g., node failures or deploying new nodes in the network topology). In this work, we focus on evaluating robustness against network load unpredictable variations. To the best of our knowledge, the present work is the first to perform such  an evaluation.

\section{Network Slice Placement Optimization Problem \label{sec:network_model}}

We present in this section the various elements composing the model for slice placement. Slices are placed on a substrate network, referred to as Physical Network Substrate (PSN) and described in Section \ref{sec::psn_model}. Slices give rise to Network Slice Placement Requests (Section \ref{sec:nspr_model}), generating a  network load defined in Section \ref{sec:network_load_modeling}. The optimization problem is formulated in Section \ref{sec:nsp_problem_statement}.

\subsection{Physical Substrate Network Modeling \label{sec::psn_model}}

\begin{figure}[hbtp]
\centering
\includegraphics[width=\linewidth]{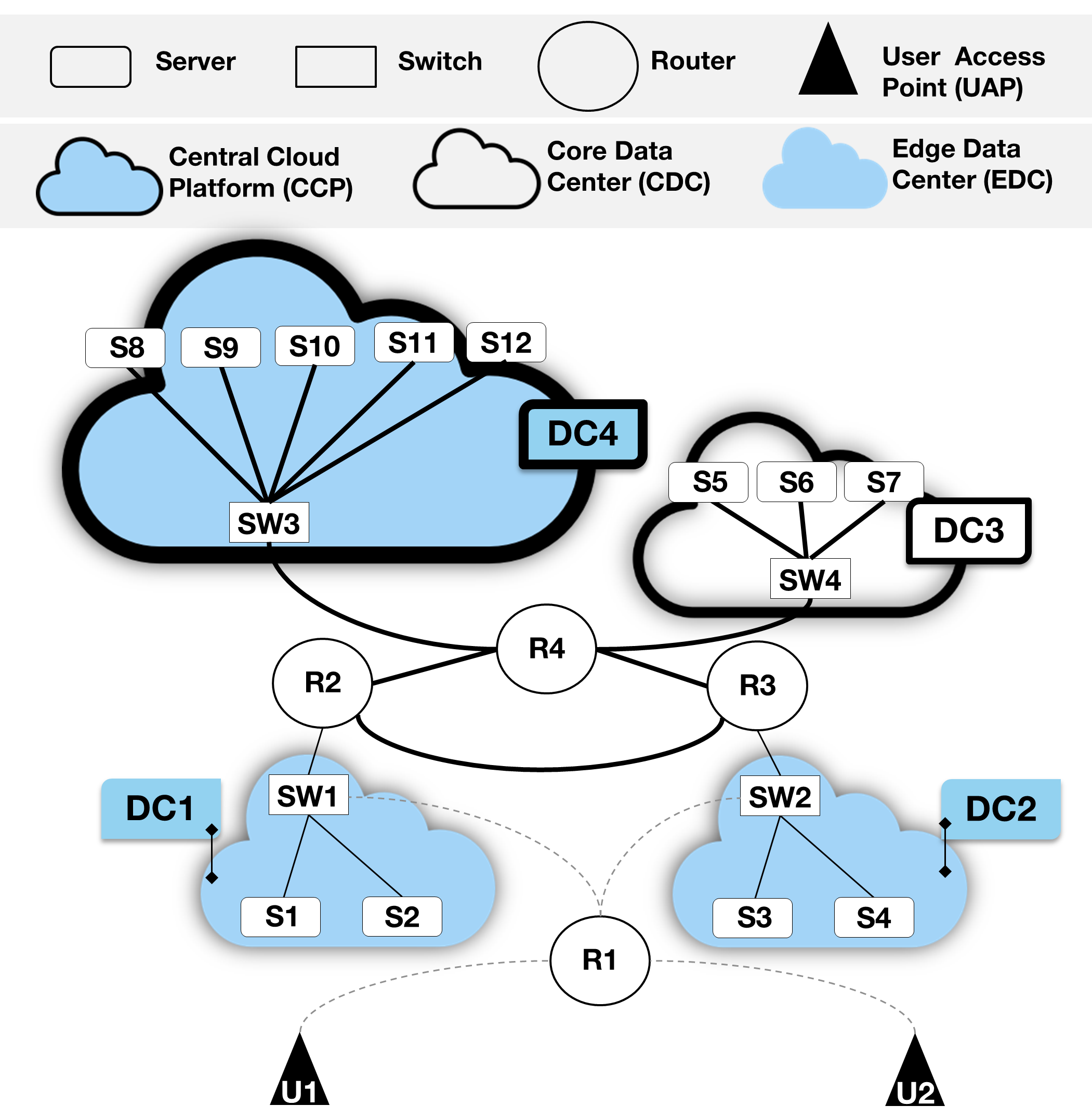}
\caption{Physical Substrate Network example.}
\label{fig:sn_model}
\end{figure}

The Physical Substrate PSN is composed of  the infrastructure resources, namely  IT resources (CPU, RAM, disk, etc.) needed for supporting  the Virtual Network Functions (VNFs) of network slices  together  with the transport network, in particular Virual Links (VLs) for interconnecting the VNFs of slices. 
As depicted in Fig.~\ref{fig:sn_model},
The PSN  is divided into three components: the Virtualized Infrastructure (VI) corresponding to IT resources, the Access Network (AN), and the Transport Network (TN). The Virtual Infrastructure (VI) hosting IT resources is  the set of Data Centers (DCs) interconnected by network elements (switches and routers). We assume that data centers are distributed in Points of Presence (PoP) or centralized (e.g., in a big cloud platform). As in  \cite{slim2018close}, we define three types of DCs with different capacities: Edge Data Centers (EDCs) close to end users but  with small resources capacities, Core Data Centers (CDCs) as regional DCs with medium resource capacities, and Central Cloud Platforms (CCPs) as national DCs with big resource capacities. We consider that slices are rooted so as to take into account the location of those users of a slice. We thus introduce an Access Network (AN) representing User Access Points (UAPs) such as Wi-Fi APs, antennas of cellular networks, etc. and Access Links. 

Users access  slices  via one UAP, which may change during the life time of a communication by a  user (e.g., because of mobility). The Transport Network (TN) is the set of routers and transmission links needed to interconnect the different DCs and the UAPs.

The complete PSN is modeled as a weighted undirected graph $G_s = (N, L)$ with parameters described in Table \ref{tab::physical_substrate_network}, where $N$ is the set of physical nodes in the PSN, and $L \subset \{(a, b) \in N \times N : a\neq b\}$ refers to a set of substrate links. Each node has a type in the set $\{$UAP, router, switch, server$\}$. The available CPU and RAM capacities on each node are defined  as $cap^{cpu}_n \in \mathbb{R}$, $cap^{ram}_n \in \mathbb{R}$ for all $n \in N$, respectively. The available bandwidth on the links are defined as $cap^{bw}_{(a,b)} \in \mathbb{R}, \forall (a,b) \in L$.
      
\begin{table}[hbtp]
\caption{PSN parameters \label{tab::physical_substrate_network}}
\begin{tabular}{@{}cc@{}}
\toprule                 
\textit{\textbf{Parameter}}                          & \textit{\textbf{Description}}             \\ \midrule
$G_s = (N,L)$       & PSN graph \\
$N$               & Network nodes  \\
$S \subset N$     & Set of servers \\
$DC$             & Set of data centers                       \\
$S_{dc} \subset S$, $\forall dc \in DC$ & Set of servers in data center $dc$        \\
 $SW_{dc}, \ \forall dc \in DC$   & Switch of of data center $dc$  \\
$L = \{(a,b) \in N \times N \wedge a \neq b\}$ & Set of physical links                     \\
$cap^{bw}_{(a,b)} \in \mathbb{R}, \forall (a,b) \in L$ & Bandwidth capacity of  link $(a,b)$ \\
$cap^{cpu}_s \in \mathbb{R}, \forall s \in S$        & available CPU capacity on server $s$                \\
$M^{cpu}_s \in \mathbb{R}, \forall s \in S$        & maximum CPU capacity of server $s$                \\
$cap^{ram}_s \in \mathbb{R}, \forall s \in S$        & available RAM capacity on server $s$ \\
$M^{ram}_s \in \mathbb{R}, \forall s \in S$        & maximum RAM capacity of server $s$ \\
$M^{bw}_s \in \mathbb{R}, \forall s \in S$        & maximum outgoing bandwidth from $s$ \\ \bottomrule
\end{tabular}
\end{table}

\subsection{Network Slice Placement Requests Modeling \label{sec:nspr_model}}

We consider that a slice is a chain of  VNFs to be placed and connected over the PSN. VNFs of a slice are grouped into a request, namely a Network Slice Placement Request (NSPR), which has to be placed on the PSN.  A  NSPR is  represented as a weighted undirected graph  $G_v = (V, E)$, with parameters described in Table~\ref{tab::nspr_parameters}, where $V$ is the set of VNFs in the NSPR, and $ E \subset \{(\bar{a}, \bar{b}) \in V \times V \wedge \bar{a} \neq \bar{b}\}$ is a set of VLs to interconnect the VNFs of the slice . The CPU and RAM requirements of each VNF of a NSPR are defined as $req^{cpu}_{v} \in \mathbb{R}$ and $req^{ram}_{v} \in \mathbb{R}$ for all $v \in V$, respectively. The bandwidth required by each VL in a NSPR is given by $req_{(\bar{a},\bar{b})}^{bw} \in \mathbb{R}$  for all  $(\bar{a},\bar{b}) \in E$.  

We consider the existence of different NSPR classes characterizing different levels of resources requirements, lifespan and arrival rate at described in Section \ref{sec:network_load_modeling}.  

\begin{table}[hbtp]
\centering
\caption{NSPR parameters \label{tab::nspr_parameters}}
\begin{tabular}{@{}cc@{}}
\toprule
\textit{\textbf{Parameter}}                                          & \textit{\textbf{Description}}                   \\ \midrule
$G_v = (V,E)$                                                                  & NSPR graph                         \\
$V$                                                                  & Set of VNFs of the NSPR                         \\
$E=\{(\bar{a},\bar{b}) \in N \times N \wedge \bar{a} \neq \bar{b}\}$ & Set of VLs of the NSPR                          \\
$req^{cpu}_{v} \in \mathbb{R}$                                         & CPU requirement of VNF $v$                      \\
$req^{ram}_{v} \in \mathbb{R}$                                         & RAM requirement of VNF $v$                      \\
$req_{(\bar{a},\bar{b})}^{bw} \in \mathbb{R}$                          & Bandwidth requirement of VL $ (\bar{a},\bar{b})$\\ \bottomrule
\end{tabular}
\end{table}


\subsection{Network Load Modeling \label{sec:network_load_modeling}}

The Network Load model allow us to control the percentage of the total network resources capacity being used at a specific instant. 

Let $J$ be the set of resources in the network (i.e., CPU, RAM, bandwidth). Let $\mathcal{K} \subset \mathbb{N}$ be the set of NSPR classes. We compute the load generated by arrivals of NSPRs of class $k \in \mathcal{K}$ for resource $j$ in $J$ as in \cite{farah2}:
\begin{equation}
    \rho^{k}_{j} = \frac{1}{C_j}\frac{\lambda^{k}}{\mu^{k}}A^{k}_{j} ,
    \label{eq:static}
\end{equation}
where $C_j$ is the total capacity of resource $j$, $A^k_j$ is the number of resource units requested by an NSPR of class $k$, $\lambda^{k}$ is the average arrival rate for an NSPR of class $k$ and $1/\mu^{k}$ is the average lifetime of an NSPR of class $k$. 

We define the global load  $\rho_{j}$  for resource $j$  as the sum 
\begin{equation}
    \rho_{j} = \sum_{k \in \mathcal{K}} \rho^{k}_{j}
    \label{eq:global}
\end{equation}
If $ 0 \leq \rho_j \leq 1$, the system is not overloaded for resource $j$; otherwise, the system is under overload conditions and the rejection of NSPRs may be high.

\subsection{Network Slice Placement Optimization Problem Statement \label{sec:nsp_problem_statement}}

The Network Slice Placement optimization problem is stated as follows: 

\begin{itemize}
    \item \textit{Given:} a NSPR graph $G_v = (V, E)$ and a PSN graph $G_s = (N, L)$,
    \item \textit{Find:} a mapping $G_v \to  \bar{G}_s =(\bar{N},\bar{L})$, $\bar{N} \subset N$, $\bar{L} \subset L$,
    \item\textit{Subject to:} the VNF CPU requirements $req^{cpu}_v, \forall v \in V$, the VNF RAM requirements $req^{ram}_v, \forall v \in V$, the VLs bandwidth requirements $req^{bw}_{(\bar{a},\bar{b})}, \forall (\bar{a},\bar{b}) \in E$, the server CPU available capacity $cap^{cpu}_s, \forall s \in S$, the server RAM available capacity $cap^{ram}_s, \forall s \in S$, the physical link bandwidth available capacity $cap^{bw}_{(a,b)}, \forall (a,b) \in L$.
    \item \textit{Objective: } maximize the network slice placement request acceptance ratio, minimize the total resource consumption and maximize load balancing.
\end{itemize}

A complete mathematical formulation of this problem can be found in \cite{HA_DRL_TNSM}.

\section{Learning framework for Network Slice Placement Optimization  \label{sec:drl_proposal}}

We describe in this section the DRL-based approach used to solve the optimization formulated in Section~\ref{sec:network_model}. As mentioned, we adopt the same approach as in \cite{HA_DRL_TNSM} but we focus here on evaluating the performance when a unpredictable network load change occurs. 
\subsection{Learning framework}
\label{sec:drl_policy}

\begin{figure}[hbtp] 
\centering
\includegraphics[width=\linewidth]{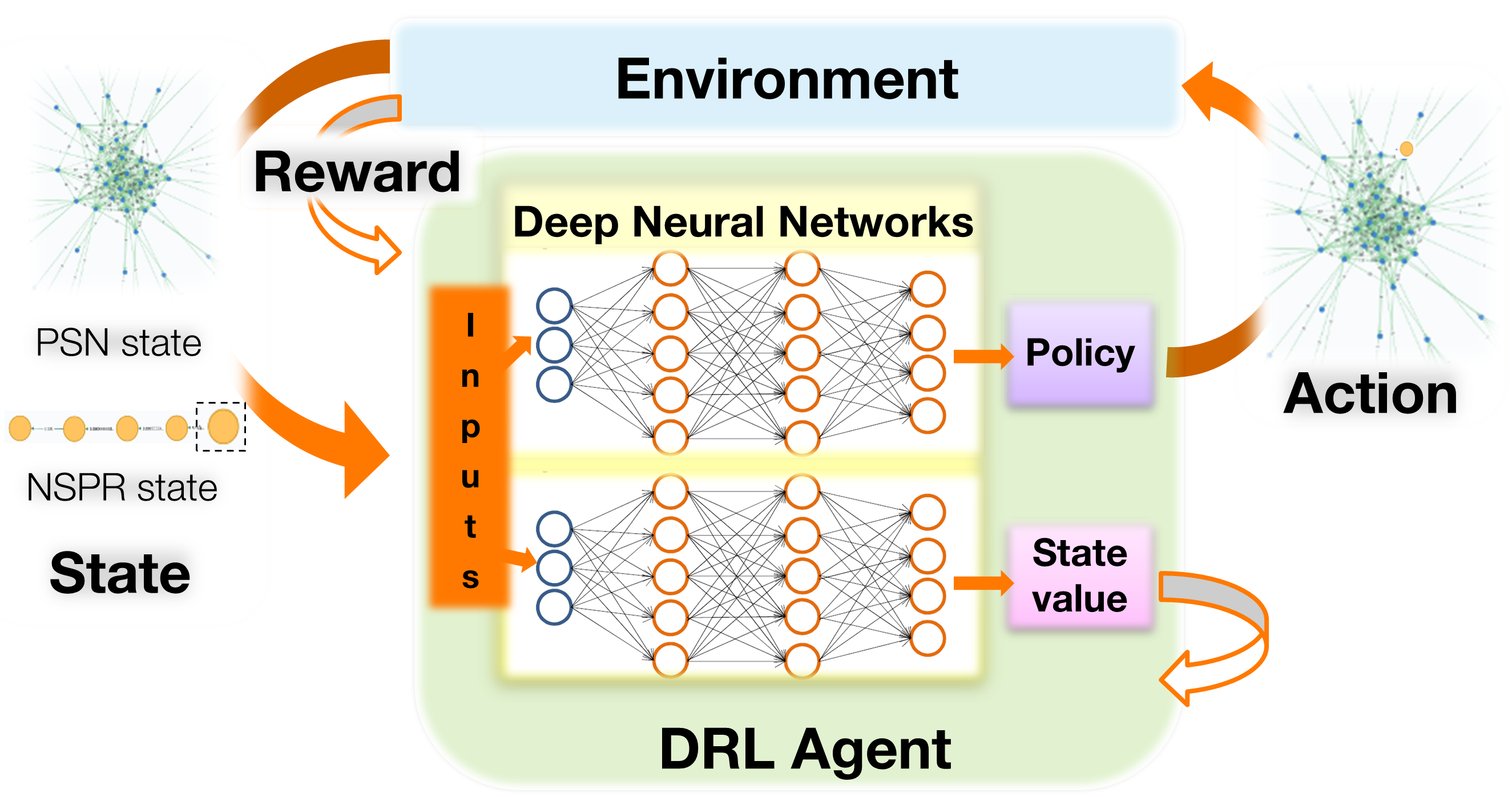}
\caption{DRL framework for Network Slice Placement Optimization} \label{fig::drl_framework_for_nsp}
\end{figure}

Fig.~\ref{fig::drl_framework_for_nsp} presents an overview of the DRL framework. The state contains the features of the PSN and NSPR to be placed. A valid action is, for a given NSPR graph $G_{v} = (V,E)$, a subgraph of the PSN graph $\bar{G_s} \subset \bar{G_s} = (N, L)$ to place the NSPR that does not violate the problem constraints described in \cite{HA_DRL_TNSM} Section \ref{sec:nsp_problem_statement}. 

The reward evaluates how good is the computed action with respect to the optimization objectives described in \cite{HA_DRL_TNSM} Section \ref{sec:nsp_problem_statement}. DNNs are trained to calculate i)  optimal actions for each state (i.e., placements with maximal rewards) and ii)  the State-value function used in the learning process.

In the following sections we describe each one of the elements of this framework. 

\subsubsection{Policy} We reuse the  framework introduced in \cite{HA_DRL_TNSM}. We denote by  $\mathcal{A}$ the set of possible actions (namely placing VNFs on nodes)  and by $\mathcal{S}$ the set of all states.  We adopt a sequential placement strategy so that we choose a node $n \in N$ where to place a specific VNF $v \in \{1,...,|V|\}$. The VNFs are sequentially placed  so that placement starts with the VNF $v=1$ and ends for the  VNF $v = |V|$.

 At each time step $t$, given a state $\sigma_t$, the learning agent select an action $a$ with probability given by  the Softmax distribution 
\begin{equation}
    \pi_{\theta}(a_{t} = a|\sigma_t) = \frac{e^{Z_{\theta}(\sigma_t,a),}}{\sum_{b \in N}e^{Z_{\theta}(\sigma_t,b)}},
    \label{eq::policy}
\end{equation}
where the  function $Z_{\theta}: \sset \times \aset \rightarrow \mathbb{R}$ yields a real value for each state and action calculated by a Deep Neural Network (DNN) as detailed in Section~\ref{sec::drl_learning}. The notation $\pi_{\theta}$ is used to indicate that policy depends on  $Z_{\theta}$. The control parameter $\theta$ represents the weights in the DNN.

\subsubsection{State representation}

As in \cite{HA_DRL_TNSM}, the \textbf{PSN state} is characterized by available server resources:   $cap^{cpu} = \{cap^{cpu}_{n}: n \in N\}$, $cap^{ram} = \{cap^{ram}_{n}: n \in N\}$ and $cap^{bw} = \{cap^{bw}_{n} = \sum_{(n,b) \in L}cap^{bw}_{(n,b)}: n \in N\}$. In addition, we keep track of the placement of the pending NSPR (i.e., the one being placed) via the vector $\chi = \{\chi_{n} \in \{0,..,|V|\} : n \in N \}$, where $\chi_{n}$ is  the number of VNFs of the current NSPR placed on node $n$.

The \textbf{NSPR state} is a view  of the current placement and is composed of four characteristics, three related to resource requirements (see Table \ref{tab::nspr_parameters} for the notation) of  the current VNF $v$ to be placed: $req^{cpu}_{v}$, $req^{ram}_{v}$ and $req^{bw}_{v} =  \sum_{(v,\bar{b}) \in E}req^{bw}_{(v,\bar{b})}$, and $m_{v} = |V| - v + 1$ the number of VNFs of the outstanding  NSPR still  to be placed.

\subsubsection{Reward function} We reuse the reward function introduced in \cite{HA_DRL_TNSM}. We precisely consider
\begin{equation}
     \small
     r_{t+1} = \left\{\begin{array}{lr}
     0, & \text{if $t < T$ and $a_{t}$ is  successful}\\
     \sum^{T}_{i=0} \delta^{a}_{i+1}\delta^{b}_{i+1}\delta^{c}_{i+1}, & \text{if $t = T$ and $a_{t}$ is successful}\\
     \delta^{a}_{t+1}, & \text{otherwise}
    \end{array}\right.
    \label{eq::reward_function}
\end{equation}
where $T$ is the number of iterations of a training episode and where the rewards  $\delta^{a}_{i+1}$, $\delta^{b}_{i+1}$, and $\delta^{c}_{i+1}$ are defined as follows:
\begin{itemize}
    \item An Action $a_t$ may lead to a successful or unsuccessful placement. We then define the Acceptance Reward value due to action $a_t$ as
\begin{equation}
     \delta^{a}_{t+1} = \left\{\begin{array}{lr}
    100, & \text{if $a_{t}$ is successful, }\\
    -100, & \text{otherwise. }
    \end{array}\right.
    \label{eq::acceptance_signal}
\end{equation}
\item  The Resource Consumption Reward value for the placement of VNF $v$ via action $a_t$ is defined by
\begin{equation}
     \delta^{c}_{t+1}= \left\{\begin{array}{lr}
    \frac{req^{bw}_{(v-1,v)}}{req^{bw}_{(v-1,v)}|P|} = \frac{1}{|P|}, & \text{if $|P|>0$, }\\
    1, & \text{otherwise. }
    \end{array}\right.
    \label{eq::resource_consumption_signal}
\end{equation}
where $P$ is the path used to place VL $(v-1,v)$. Note that a maximum  $\delta^{c}_{t+1} = 1$ is given when $|P|=0$, that is, when VNFs $v-1$ and $v$ are placed on the same server.
\item  The Load Balancing Reward value for the placement of VNF $v$ via $a_t$
\begin{equation}
    \delta^{b}_{t+1} = \frac{cap^{cpu}_{a_t}}{M^{cpu}_{a_{t}}} + \frac{cap^{ram}_{a_t}}{M^{ram}_{a_{t}}}.
    \label{eq::load_balancing_signal}
\end{equation}
\end{itemize}

\subsection{Adaptation of DRL and Introduction of a Heuristic Function
\label{sec:aidedDRL}}

\subsubsection{Proposed Deep Reinforcement Learning Algorithm \label{sec::drl_learning}}

As in \cite{HA_DRL_TNSM}, we use a single thread version of the A3C Algorithm introduced in \cite{a3c}. 
This algorithm relies on two DNNs that are trained in parallel: i) the Actor Network with the parameter $\theta$, which is used to generate the policy $\pi_{\theta}$ at each time step, and ii) the Critic Network with the parameter $\theta_{v}$ which generates an estimate $\nu^{\pi_{\theta}}_{\theta_{v}}(\sigma_t)$ for the  State-value function defined by $$\nu_{\pi}(t|\sigma)=\mathbb{E}_{\pi}\left[\sum^{T-t-1}_{k=0}\gamma^{k} r_{t+k+1} | \sigma_t = \sigma \right],$$
for some discount parameter $\gamma$. 

\begin{figure}[hbtp] 
\centering
\includegraphics[width=\linewidth]{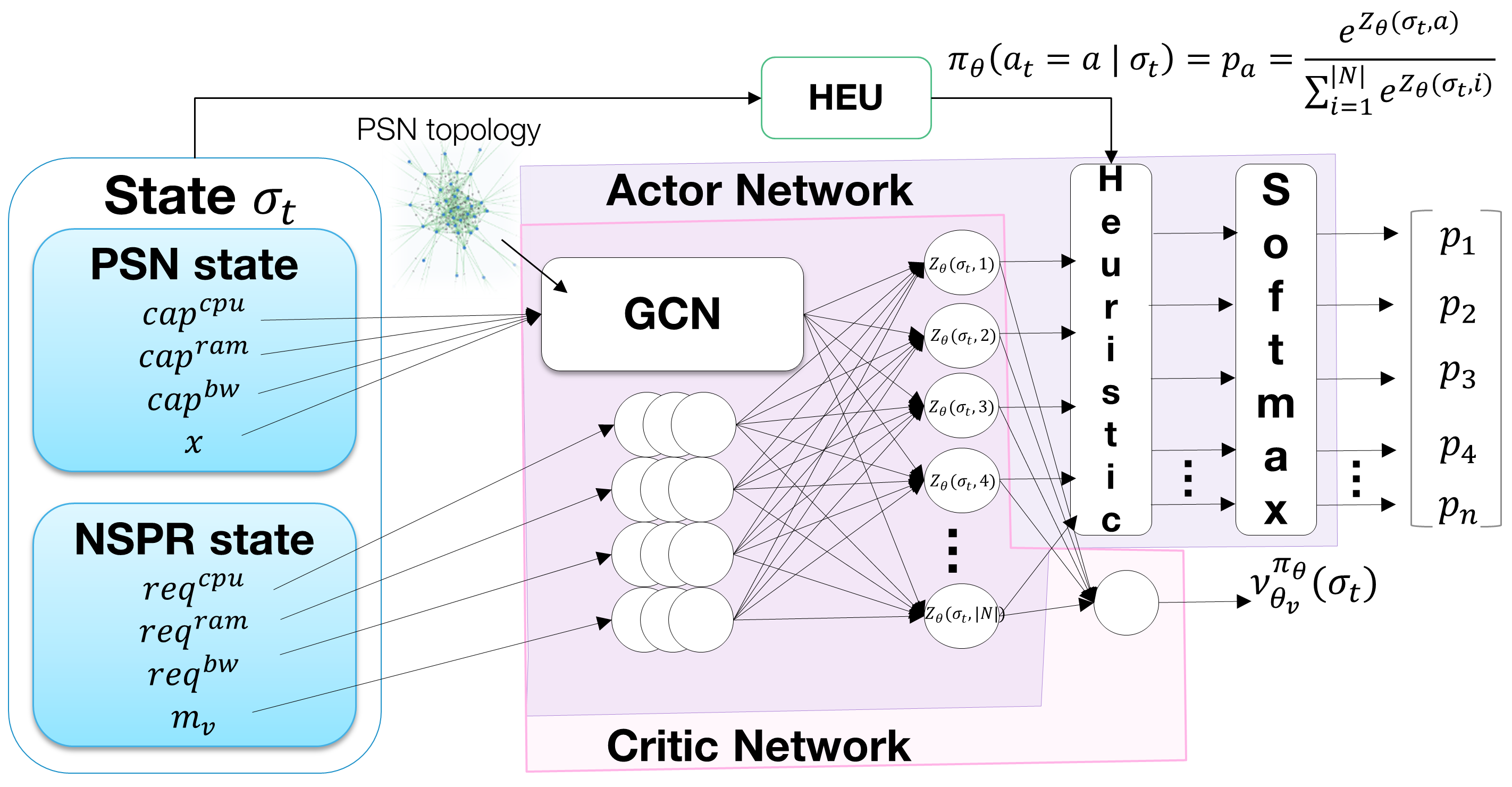}
\caption{Reference framework for the proposed learning algorithms.} \label{fig::ha_advantage_actor_critic_architecture}
\end{figure}

As depicted in Fig.~\ref{fig::ha_advantage_actor_critic_architecture} both Actor and Critic Networks have almost identical structure. As in \cite{p1}, we use the GCN formulation proposed by \cite{kipf_gcn} to automatically extract advanced characteristics of the PSN. The characteristics produced by the GCN represent semantics of the PSN topology by encoding and accumulating characteristics of neighbour nodes in the PSN graph. The size of the neighbourhood is defined by the order-index parameter $K$. 

As in \cite{p1}, we consider in the following $K=3$ and perform automatic extraction of 60 characteristics per PSN node. 

The NSPR state characteristics are separately transmitted to a fully connected layer with 4 units. The characteristics extracted by both layers and the GCN layer are combined into a single column vector of size $60|N| + 4$ and passed through a fully connected layer with $|N|$ units.

In the Critic Network, the outputs are forwarded to a single neuron, which is used to calculate the state-value function estimation $\nu^{\pi_{\theta}}_{\theta_{v}}(\sigma_t)$. In the Actor Network, the outputs represent the values of the function $Z_{\theta}$ introduced in Section \ref{sec:drl_policy}. These values are injected into a Softmax layer that transforms them into a Softmax distribution that corresponds to the policy $\pi_{\theta}$.

During the training phase, at each time step $t$, the  A3C algorithm uses the Actor Network to calculate the policy $\pi_{\theta}(.|\sigma_t)$. An action $a_t$ is sampled using the policy and performed on the environment. The Critic Network is used to calculate the state-value function approximation $\nu^{\pi_{\theta}}_{\theta_{v}}(\sigma_t)$. The learning agent receives then the reward $r_{t+1}$ and next state $\sigma_{t+1}$ from the environment and the placement process continues until a terminal state is reached, that is, until the Actor Network returns an unsuccessful action or until the current NSPR is fully placed. At the end of the training episode, the A3C algorithm updates parameters $\theta$ and $\theta_{v}$ by using the same rules as in \cite{HA_DRL_TNSM}.

\subsubsection{Introduction of a Heuristic Function}
\label{heuristicfunc}
To guide  the learning process, we use as in \cite{HA_DRL_TNSM} the placement heuristic introduced in  \cite{cnsm_2020}. This yields the  HA-DRL algorithm. More precisely, from the reference framework shown in Fig.~\ref{fig::ha_advantage_actor_critic_architecture},  we proposed to include in the  Actor Network  the Heuristic layer that calculates an Heuristic Function $H: \sset \times \aset \rightarrow \mathbb{R}$ based on external information provided by the heuristic method,  referred as HEU. 

Let $Z_{\theta}$ be the function computed by the fully connected layer of the Actor Network that maps each state and action to a real value which is after converted by the Softmax layer into the selection probability of the respective action (see Section \ref{sec:drl_policy}).

Let $\bar{a}_{t} = \text{argmax}_{a \in \aset}\,Z_{\theta}(\sigma_t,a)$ be the action with the highest $Z_{\theta}$ value for state $\sigma_{t}$. Let $a^{*}_{t}=HEU(\sigma_t)$ be the action derived by the HEU method at time step $t$ and the preferred action to be chosen. $H(\sigma_t,a^{*}_t)$ is shaped to allow the value of $Z_{\theta}(\sigma_t,a^{*}_t)$ to become closer to the value of $Z_{\theta}(\sigma_t,\bar{a}_t)$. The aim is to turn $a^{*}_t$ into one of the likeliest actions to be chosen by the policy.

The Heuristic Function is then formulated as
\begin{multline}
     \small
     H(\sigma_t,a_t) =   \left\{\begin{array}{lr}
     Z_{\theta}(\sigma_t,\bar{a}_{t}) -  Z_{\theta}(\sigma_t,a_t) + \eta, & \text{if $a_{t}=a^{*}_{t}$}\\
     0, & \text{otherwise}
    \end{array}\right.
    \label{eq::heuristic_function}
\end{multline}
where $\eta$ parameter is a small real number. 

During the training process the Heuristic layer calculates $H(\sigma_t,.)$ and updates the $Z_{\theta}(\sigma_t,.)$ values by using the following equation:
\begin{equation}
    Z_{\theta}(\sigma_t,.) = Z_{\theta}(\sigma_t,.) + \xi H(\sigma_t,.)^{\beta} \label{eq:z_update} 
\end{equation}
The Softmax layer then computes the policy using the modified $Z_{\theta}$. Note the action returned by $a^{*}_{t}$ will have a higher probability to be chosen. The $\xi$ and $\beta$ parameters are used to control how much HEU influence the policy.

\subsection{Implementation Remarks}

All resource-related characteristics are normalized to be in  $[0,1]$. This is done by dividing $cap^{j}$ and $req^{j}$, $j \in \{$cpu, ram,bw$\}$, by $\max_{n \in N}M^{j}_{n}$. With regard to the DNNs, we have implemented the Actor and Critic as two independent Neural Networks. Each neuron has a bias assigned. We have used the hyperbolic tangent (tanh) activation for non-output layers of the Actor Network and Rectified Linear Unit (ReLU) activation for all layers of the Critic Network. We have normalized positive global rewards to be in  $[0,10]$. During the training phase, we have considered the policy as a Categorical distribution and used it to sample the actions randomly.
\section{Implementation and  Evaluation Results \label{sec:evaluation}}

\subsection{Implementation Details \& Simulator Settings}

\subsubsection{Experimental setting} We developed a simulator in Python containing: i) the elements of the Network Slice Placement Optimization problem (i.e., PSN and NSPR); ii) the DRL and HA-DRL algorithms. We used the PyTorch framework to implement the DNNs. Experiments were run in a 2x6 cores @2.95Ghz 96GB machine.

\subsubsection{Physical Substrate Network Settings} \label{sec::substrate_network_settings}
We consider a PSN that could reflect the infrastructure of an operator as discussed in \cite{farah2}. In this network, three types of DCs are introduced as in Section~\ref{sec:network_model}. Each CDC is connected to three EDCs which are distant of 100 km. CDCs are interconnected and connected to one CCP that is 300 km away.

We consider 15 EDCs each one with 4 servers, 5 CDCs each with 10 servers and 1 CCP with 16 servers. The CPU and RAM capacities of each server are 50 and 300 units, respectively. A bandwidth capacity of 100 Gbps is given to intra-data center links inside CDCs and CCP, 10Gbps being the bandwidth for intra-data center links inside EDCs. Transport links connected to EDCs have 10Gpbs of bandwidth capacity. Transport links between CDCs have 100Gpbs of bandwidth capacity as well for the ones between CDCs and the CCP. 

\subsubsection{Network Slice Placement Requests Settings \label{sec::network_slice_placement_requests_settings}}

We consider NSPRs to have the Enhanced Mobile Broadband (eMBB) setting described in \cite{cnsm_2020}. Each NSPR is composed of 5 or 10 VNFs (see Section \ref{sec:network_loads}). Each VNF requires 25 units of CPU and 150 units of RAM. Each VL requires 2 Gbps of bandwidth.

\subsection{Algorithms \& Experimental Setup }\label{sec:algorithms_tested}
\subsubsection{Training Process \& Hyper-parameters}
We consider a training process with maximum duration of 6 hours for the considered algorithms. We perform seven independent runs of each algorithm to assess their average performance in terms of the metrics introduced below (see Section \ref{sec:ev_metrics}). 

After performing Hyper-parameter search, we set the learning rates for the Actor and Critic networks of DRL and HA-DRL algorithms to $\alpha = 5 \times 10^{-5}$ and $\alpha' = 1.25 \times 10^{-3}$, respectively.

We program four versions of HA-DRL agents, each with a different value for the $\beta$ parameter of the heuristic function formulation (see Section \ref{heuristicfunc}). We set in addition the parameters $\xi = 1$ and $\eta = 0$.

\subsubsection{Network load calculation}\label{sec:network_loads}
Network loads are calculated using CPU resource but the analysis could easily be applied to RAM; we use the network load model introduced in Section \ref{sec:network_loads}. We consider two NSPR classes: i) a Volatile class and ii) a Long term class. 

The differences between the two classes are related to their resource requirements and their lifespans as Volatile requests have 5 VNFs and a life-span of 20 simulation time units and Long-term requests have 10 VNFs and a life span of 500 simulation time units. 

\subsubsection{Network load change scenarios}\label{sec:network_load_disruption}
We consider that the network runs in a standard regime under  a network load being equal to 40\% (i.e., $\rho=0.4$) and that the NSPRs of each class generate half of the total load. 

In each experiment, the learning agent is trained during approximately 4 hours for this network load regime. Then a stair-stepped network load change occurs. We simulated eight different network load change levels. Each network load change level is characterized by the addition of a certain amount of extra network load ranging from 10\% to 80\% (causing system overload).

\subsection{Evaluation Metrics \label{sec:ev_metrics}}
To characterize the performance of the placement algorithms, we consider one performance metric called Acceptance Ratio per Training phase (TAR). This metric represents the Acceptance Ratio obtained in each training phase, i.e., each part of the training process, corresponding to $500$ NSPR arrivals or $500$ episodes. It is calculated as follows: $\frac{\mathrm{\# accepted \; NSPRs}}{500}$. This metric allows us to better observe the evolution of algorithm performance over time since it measures algorithm performance in independent parts (phases) of the training process without accumulating the performance of previous training phases.

Based on this metric, we identify three other important metrics used in our results discussion: 
\begin{enumerate}
    \item \textbf{Rupture TAR:} it is the TAR obtained in the training phase where the network load change occurs, i.e., the rupture phase;
    \item \textbf{Last TAR:} it is the TAR obtained in the training phase that  is prior to the rupture phase;
    \item \textbf{Average TAR:} it is the average of the TARs obtained in the 30 phases preceding the rupture phase;
    \item \textbf{TAR standard deviation:} it is the standard deviation of the TARs obtained in the 30 phases preceding the rupture phase; 
\end{enumerate}

\subsection{ Evaluation of the impact of network load change}
\begin{figure}[hbtp]
\centering
\begin{subfloat}[Addition of 10\% of network load.]
 {\includegraphics[width=.85\linewidth]{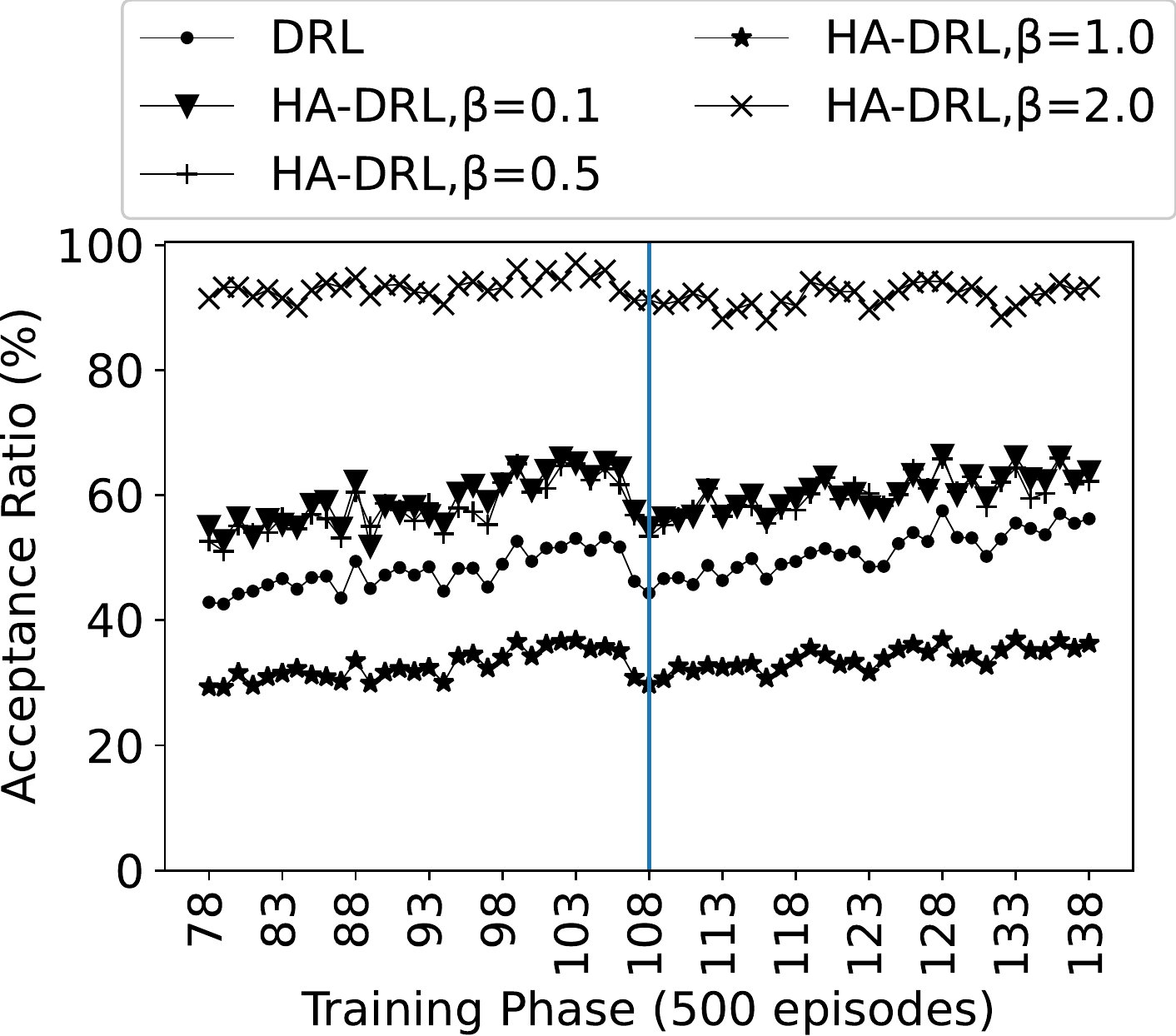}\label{fig:disruption_10}}
\end{subfloat}
\begin{subfloat}[Addition of 20\% of network load.]
 {\includegraphics[width=.85\linewidth]{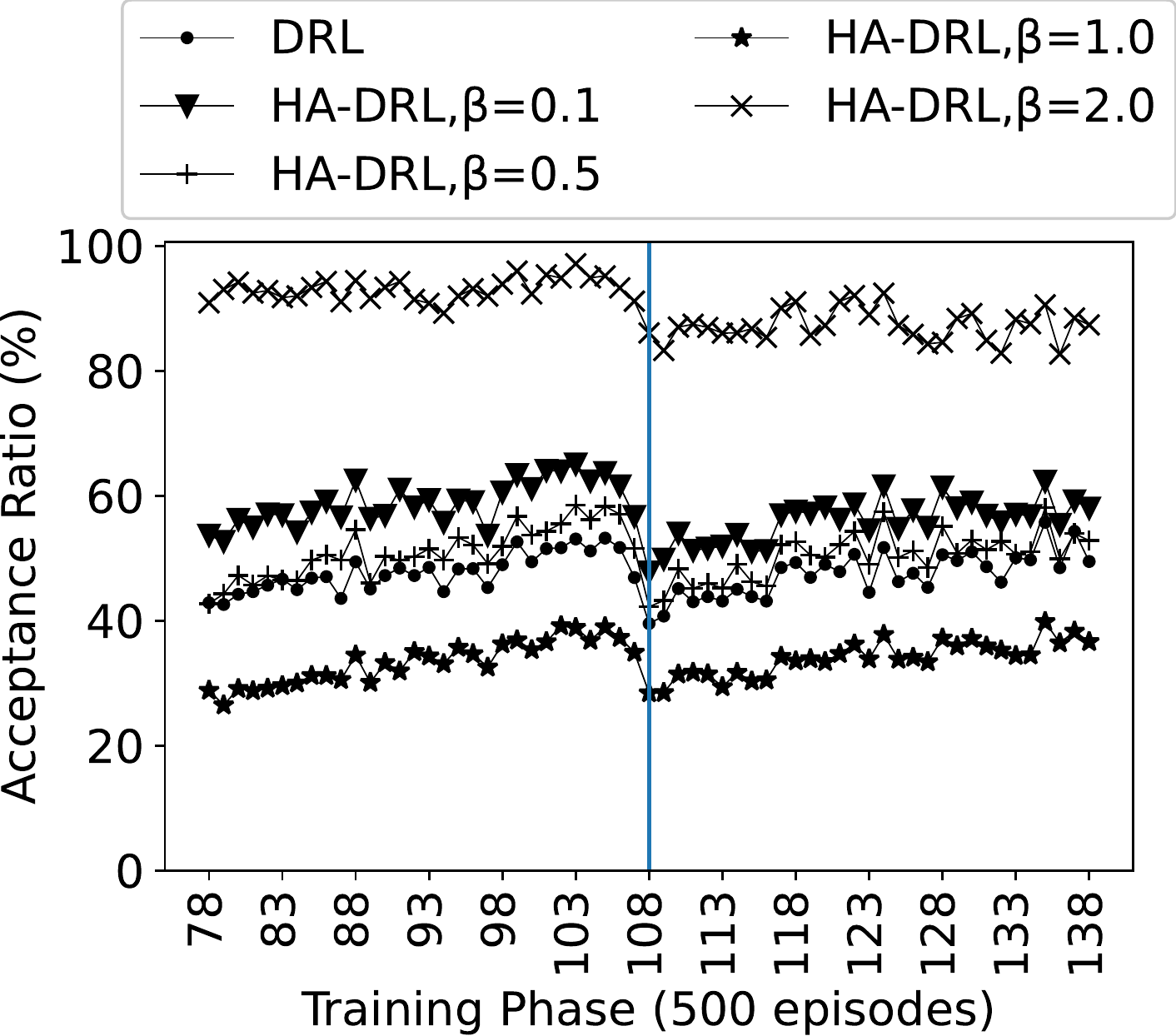}\label{fig:disruption_20}}
\end{subfloat}
\begin{subfloat}[Addition of 30\% of network load.]
 {\includegraphics[width=.85\linewidth]{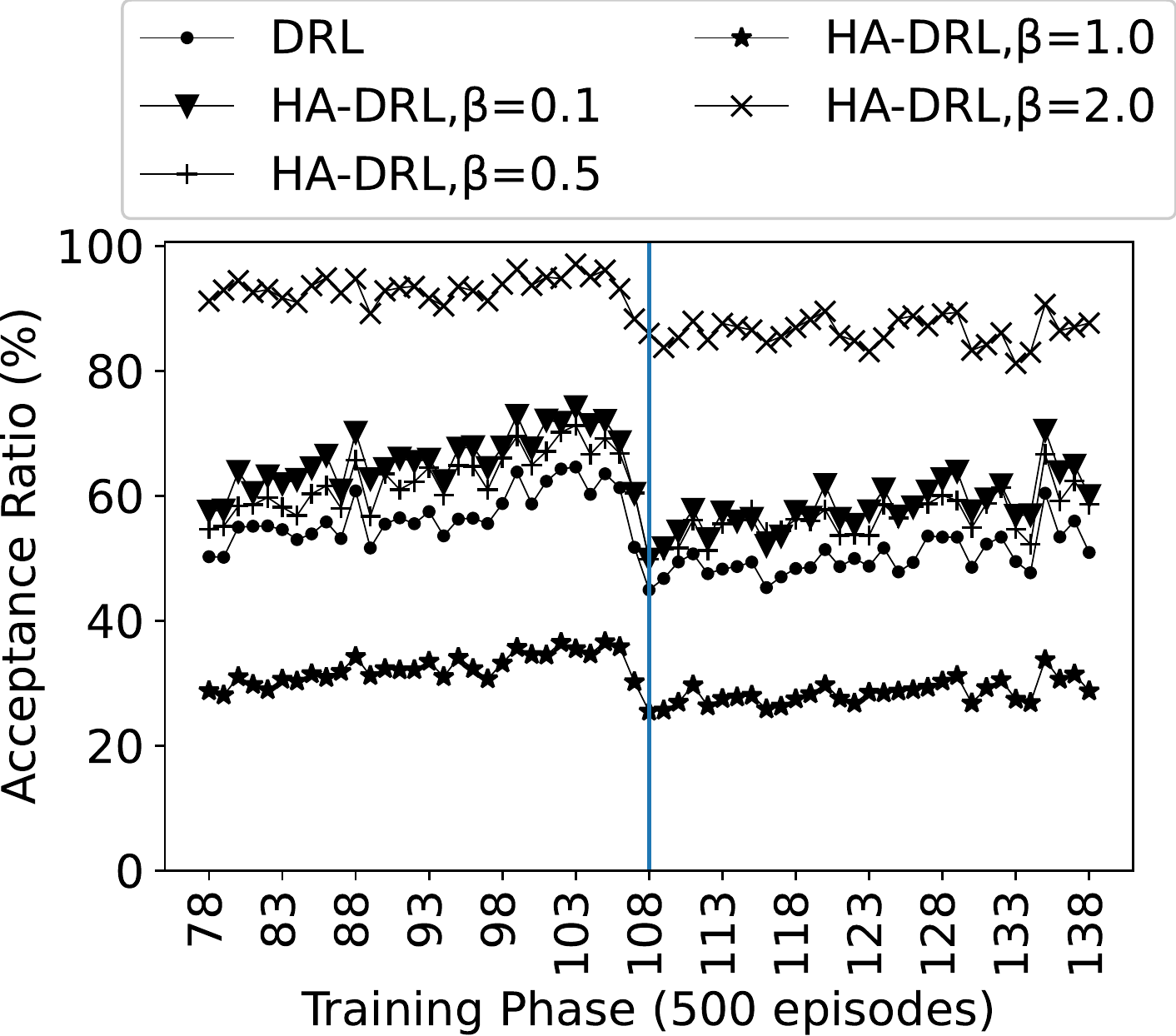}\label{fig:disruption_30}}
\end{subfloat}
\caption{Evaluation of impact of network load disruption on TAR: under-loaded scenarios \label{fig:disruption_levels_1}}
\end{figure}

\begin{figure}[hbtp]
\centering
\begin{subfloat}[Addition of 40\% of network load.]
 {\includegraphics[width=.85\linewidth]{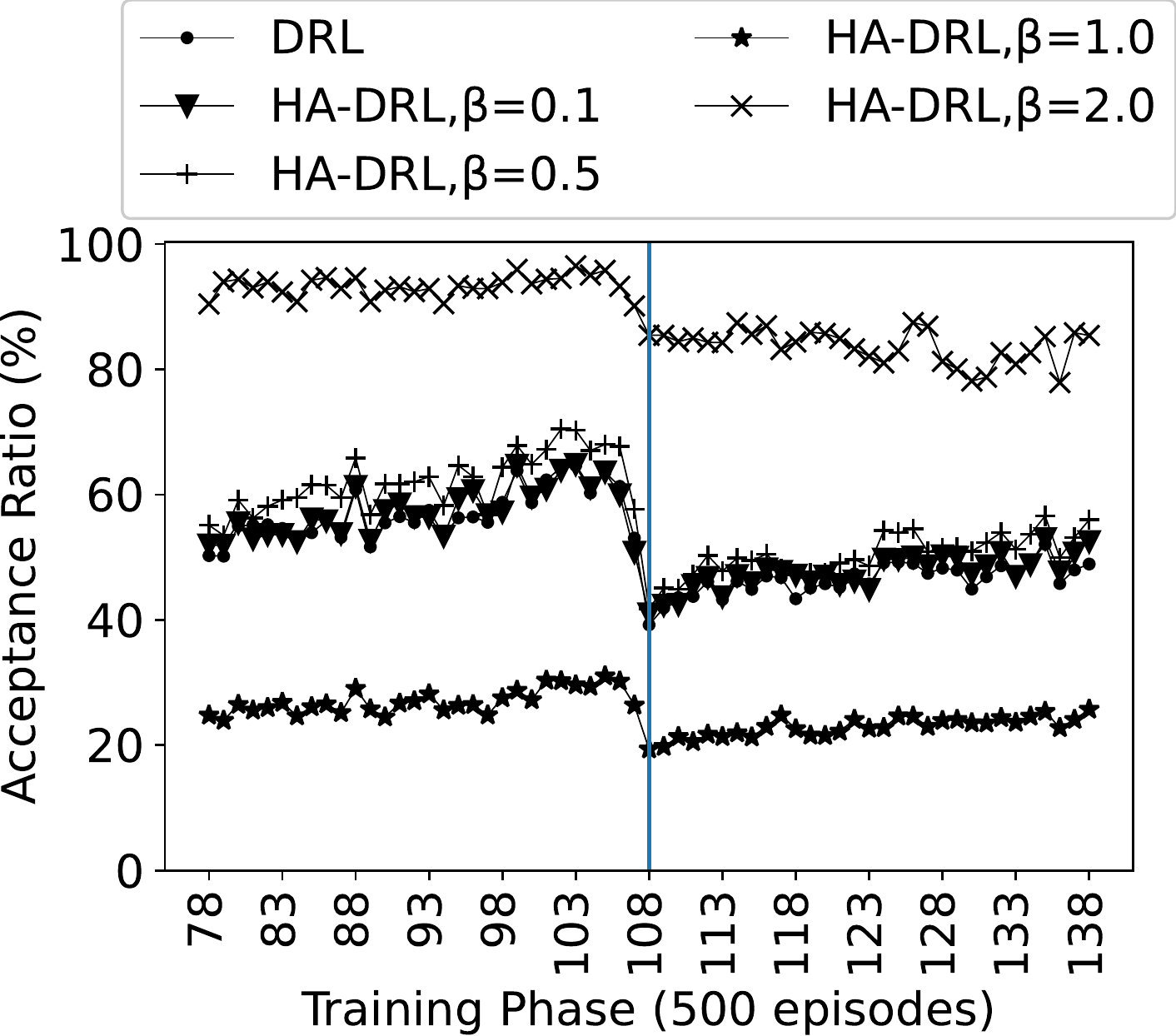}\label{fig:disruption_40}}
\end{subfloat}
\begin{subfloat}[Addition of 50\% of network load.]
 {\includegraphics[width=.85\linewidth]{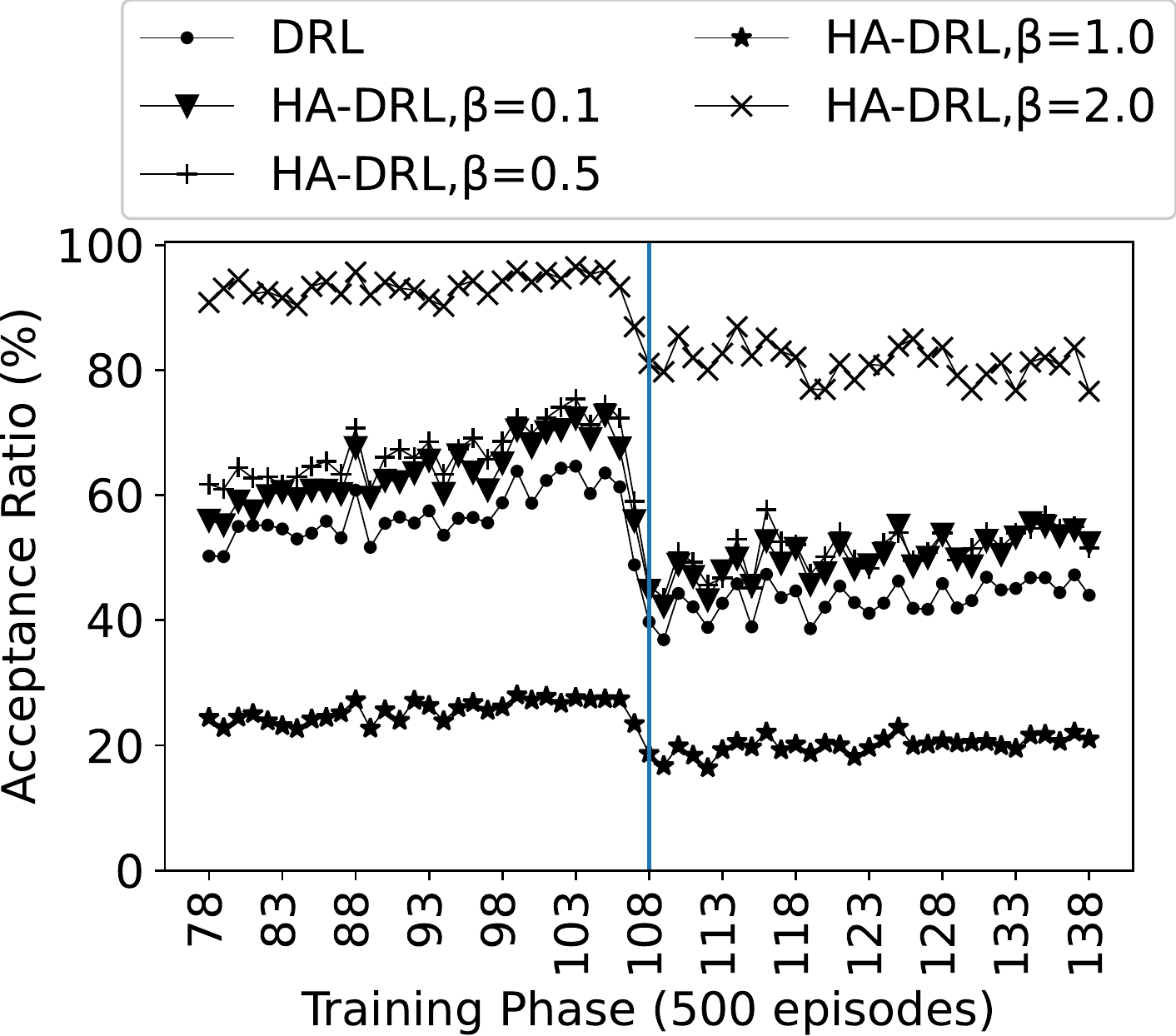}\label{fig:disruption_50}}
\end{subfloat}
\begin{subfloat}[Addition of 60\% of network load.]
 {\includegraphics[width=.85\linewidth]{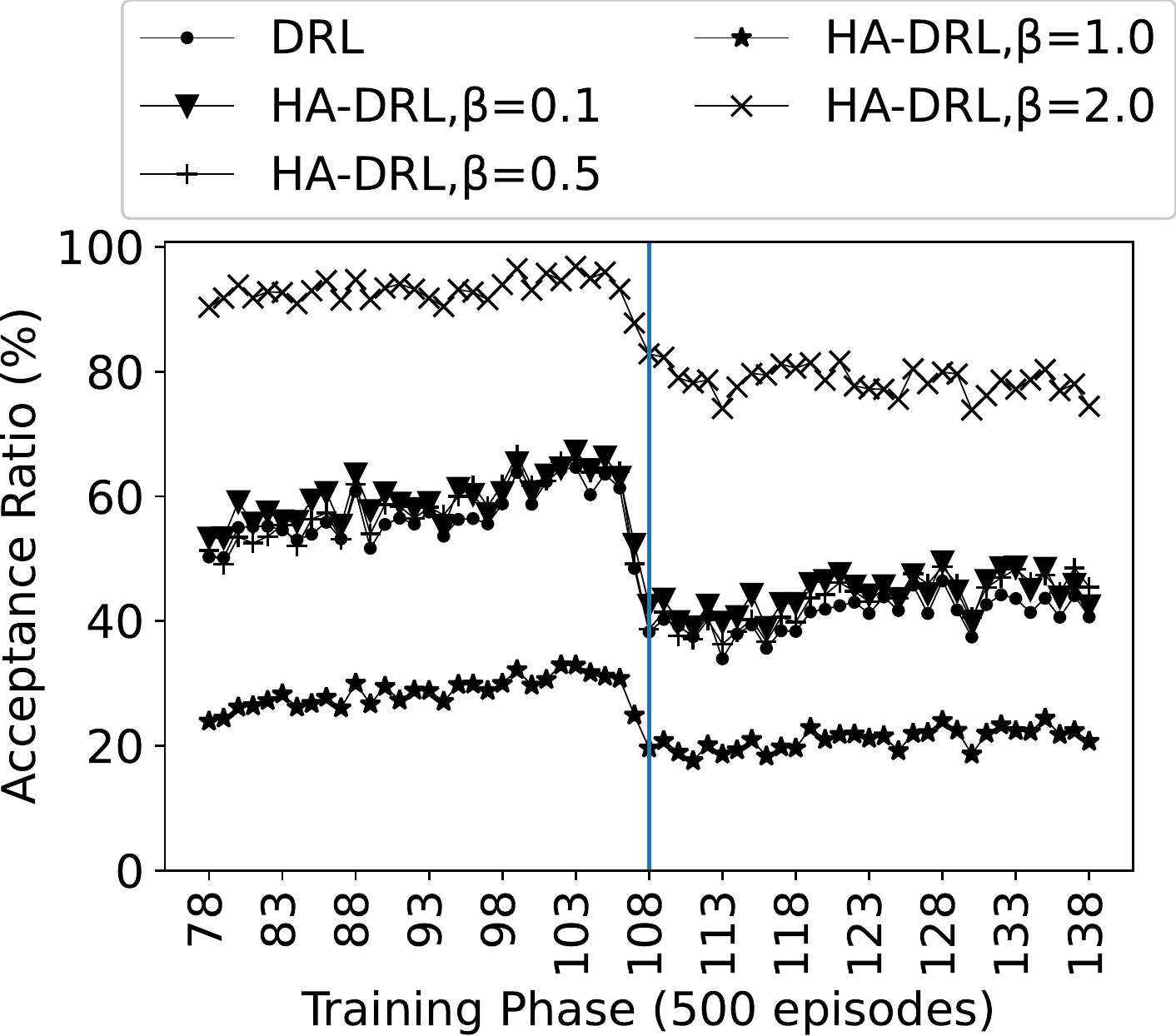}\label{fig:disruption_60}}
\end{subfloat}
\caption{Evaluation of impact of network load disruption on TAR: critical scenarios\label{fig:disruption_levels_2}}
\end{figure}

\begin{figure}[hbtp]
\centering
\begin{subfloat}[Addition of 70\% of network load.]
 {\includegraphics[width=.87\linewidth]{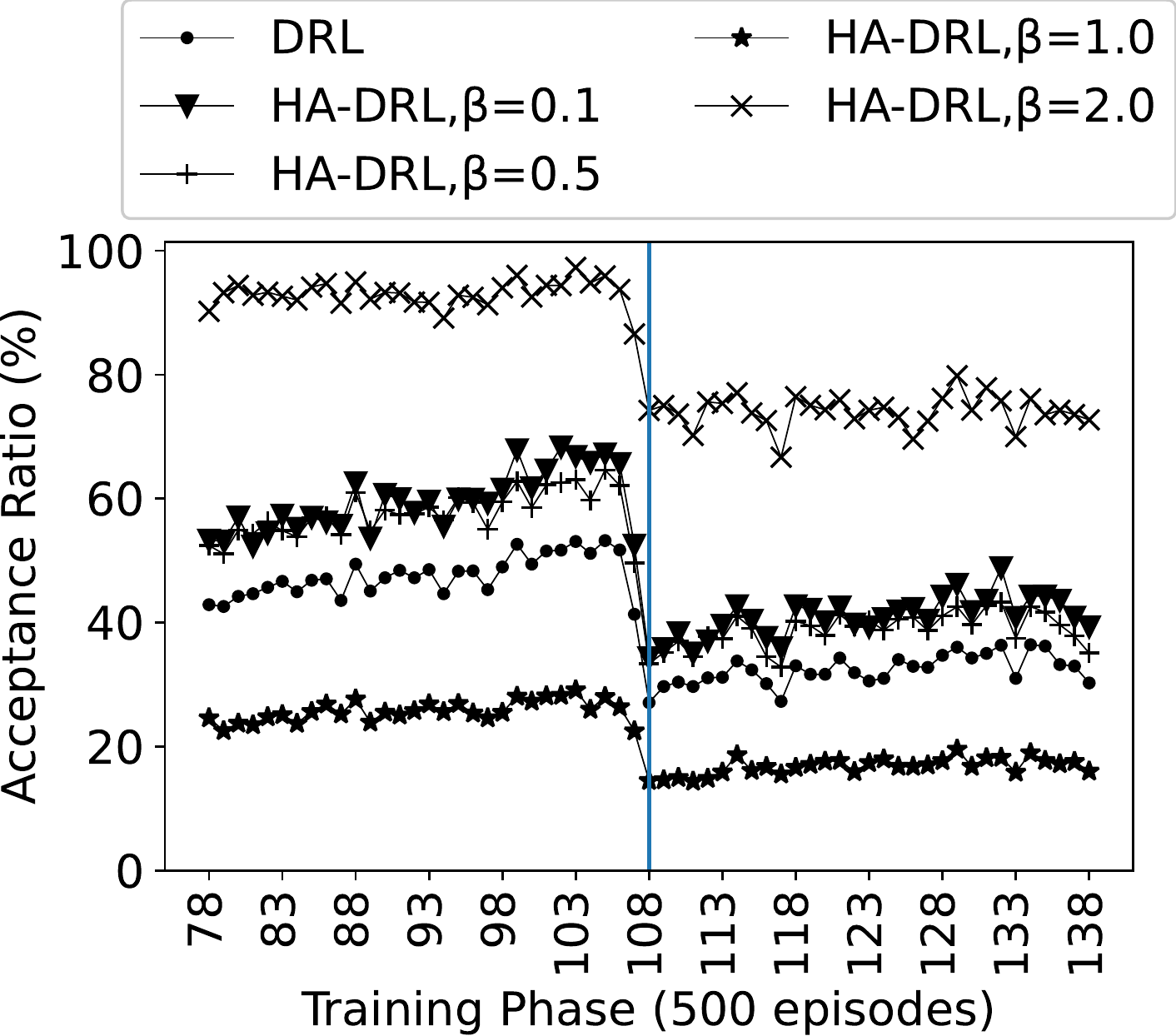}\label{fig:disruption_70}}
\end{subfloat}
\begin{subfloat}[Addition of 80\% of network load.]
 {\includegraphics[width=.87\linewidth]{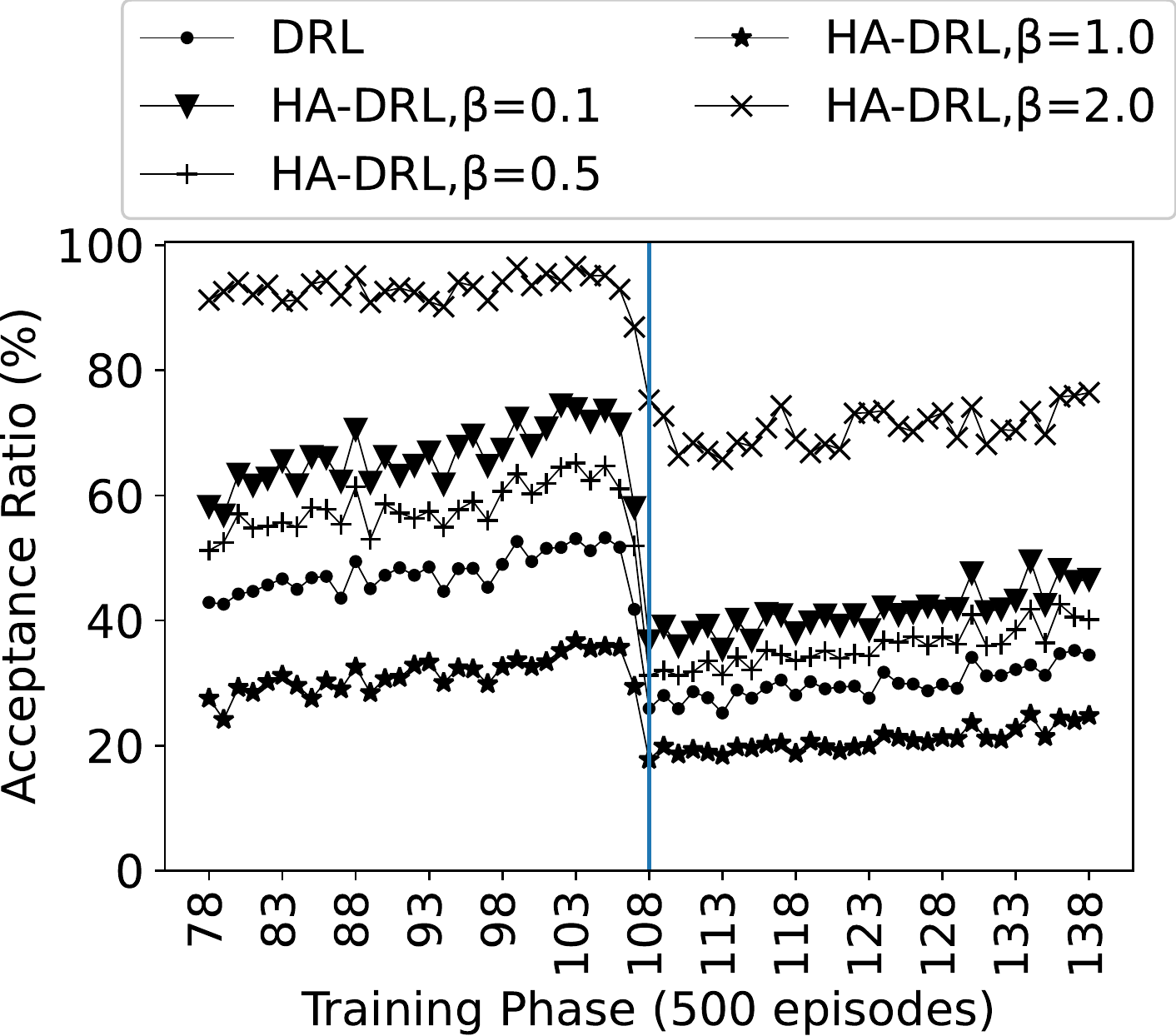}\label{fig:disruption_80}}
\end{subfloat}
\caption{Evaluation of impact of network load disruption on TAR: overloaded scenarios \label{fig:disruption_levels_3}}
\end{figure}

Fig.~\ref{fig:disruption_levels_1}, \ref{fig:disruption_levels_2} and \ref{fig:disruption_levels_3}
capture the impact of different network load change levels on the TARs obtained by the different evaluated algorithms. The rupture phase is identified by a blue vertical line in the various figures.

We can observe in Fig.~\ref{fig:disruption_levels_1}, \ref{fig:disruption_levels_2} and \ref{fig:disruption_levels_3} that with the reduced training time of 6 hours the only algorithm that has near optimal performance after 108 training phases is HA-DRL, with $\beta=2.0$. This is due to the fact that the strong influence of the  Heuristic Function helps the algorithm to become stable more quickly as  discussed in \cite{HA_DRL_TNSM} and\cite{cnsm_2021}.

We can also observe by the shape of the different curves in Fig.~\ref{fig:disruption_levels_1}, \ref{fig:disruption_levels_2}, and \ref{fig:disruption_levels_3} that, as expected, all the algorithms have some variability in their performance during the training phases. In addition, these figures show that the performance of all the algorithms is affected at various levels by the network load change and that, generally speaking, the higher the amount of extra network load added, the lower is the TAR after the change. 

Finally, we can also see that the only algorithm to keep a near optimal performance even in overloaded scenarios shown in  Fig.~\ref{fig:disruption_levels_3} is HA-DRL, with $\beta=2.0$. 

Tables \ref{tab:drl_results}, \ref{tab:ha-drl_0.1_results},\ref{tab:ha-drl_0.5_results}, \ref{tab:ha-drl_1.0_results}, and \ref{tab:ha-drl_2.0_results} present other performance metrics related to the various evaluated algorithms. The columns "Rupture TAR - Avg. TAR" and "Rupture TAR - Last TAR" indicate how much the performance of the algorithms drops in the rupture phase when compared with the Average TAR and Last TAR, respectively.

The TAR Standard Deviation column indicates the TAR Standard Deviation metric described in Section \ref{sec:ev_metrics}. 

\begin{table}[hbtp]
\centering
\caption{DRL algorithm results}
\label{tab:drl_results}
\begin{tabularx}{\linewidth}{@{}cLLL@{}}
\toprule
\begin{tabular}[c]{@{}c@{}}Network Load\\ Disruption Level (\%)\end{tabular} & Rupture TAR - Avg. TAR (\%) & Rupture TAR - Last TAR (\%) & TAR Standard Deviation (\%) \\ \midrule
+10 & -3.37  & -1.89  & 3.10 \\
+20 & -8.19  & -7.37  & 3.09 \\
+30 & -11.89 & -6.83  & 4.17 \\
+40 & -17.68 & -13.8  & 4.12 \\
+50 & -17.00 & -9.11  & 4.32 \\
+60 & -18.50 & -10.20 & 4.35 \\
+70 & -20.46 & -14.26 & 3.30 \\
+80 & -21.65 & -15.86 & 3.27 \\ \bottomrule
\end{tabularx}
\end{table}

\begin{table}[hbtp]
\centering
\caption{HA-DRL, $\beta=0.1$ algorithm results}
\label{tab:ha-drl_0.1_results}
\begin{tabularx}{\linewidth}{@{}cLLL@{}}
\toprule
\begin{tabular}[c]{@{}c@{}}Network Load\\Disruption Level (\%)\end{tabular} & Rupture TAR - Avg. TAR (\%) & Rupture TAR - Last TAR (\%) & TAR Standard Deviation (\%) \\ \midrule
+10 & -4.13  & -2.60  & 4.07 \\
+20 & -11.02 & -8.91  & 3.51 \\
+30 & -16.00 & -10.54 & 4.50 \\
+40 & -16.28 & -9.83  & 4.13 \\
+50 & -18.66 & -11.14 & 5.05 \\
+60 & -17.02 & -9.80  & 3.99 \\
+70 & -25.13 & -18.20 & 4.93 \\
+80 & -29.41 & -21.31 & 4.85 \\ \bottomrule
\end{tabularx}
\end{table}

\begin{table}[ht]
\centering
\caption{HA-DRL, $\beta=0.5$ algorithm results}
\label{tab:ha-drl_0.5_results}
\begin{tabularx}{\linewidth}{@{}cLLL@{}}
\toprule
\begin{tabular}[c]{@{}c@{}}Network Load\\ Disruption Level (\%)\end{tabular} & Rupture TAR - Avg. TAR (\%) & Rupture TAR - Last TAR (\%) & TAR Standard Deviation (\%) \\ \midrule
+10 & -4.55  & -3.43  & 3.95 \\
+20 & -8.80  & -9.37  & 4.21 \\
+30 & -12.78 & -10.66 & 4.59 \\
+40 & -20.33 & -15.94 & 4.61 \\
+50 & -21.24 & -13.43 & 4.56 \\
+60 & -19.46 & -10.46 & 5.08 \\
+70 & -24.28 & -16.26 & 3.75 \\
+80 & -26.78 & -20.71 & 3.88 \\ \bottomrule
\end{tabularx}
\end{table}

\begin{table}[t]
\centering
\caption{HA-DRL, $\beta=1.0$ algorithm results}
\label{tab:ha-drl_1.0_results}
\begin{tabularx}{\linewidth}{@{}cLLL@{}}
\toprule
\begin{tabular}[c]{@{}c@{}}Network Load\\ Disruption Level (\%)\end{tabular} & Rupture TAR - Avg. TAR (\%) & Rupture TAR - Last TAR (\%) & TAR Standard Deviation (\%) \\ \midrule
+10 & -2.96  & -1.11  & 2.37 \\
+20 & -4.94  & -6.49  & 3.50 \\
+30 & -6.93  & -4.71  & 2.37 \\
+40 & -7.67  & -7.00  & 1.97 \\
+50 & -6.80  & -4.77  & 1.72 \\
+60 & -8.95  & -5.29  & 2.45 \\
+70 & -11.25 & -8.00  & 1.73 \\
+80 & -13.62 & -11.69 & 2.89 \\ \bottomrule
\end{tabularx}
\end{table}

\begin{table}[ht]
\centering
\caption{HA-DRL, $\beta=2.0$ algorithm results}
\label{tab:ha-drl_2.0_results}
\begin{tabularx}{\linewidth}{@{}cLLL@{}}
\toprule
\begin{tabular}[c]{@{}c@{}}Network Load\\ Disruption Level (\%)\end{tabular} &Rupture TAR - Avg. TAR (\%) & Rupture TAR - Last TAR (\%) & TAR Standard Deviation (\%) \\ \midrule
+10 & -2.04  & 0.09   & 2.37 \\
+20 & -7.01  & -5.09  & 3.50 \\
+30 & -7.15  & -2.31  & 2.37 \\
+40 & -7.90  & -4.69  & 1.97 \\
+50 & -12.13 & -5.86  & 1.72 \\
+60 & -10.24 & -4.94  & 2.45 \\
+70 & -18.83 & -12.34 & 1.73 \\
+80 & -17.79 & -11.69 & 2.89 \\ \bottomrule
\end{tabularx}
\end{table}

Those tables confirm that in general the performance gaps, i.e., the gaps between the Rupture TAR and Average or Last TAR, grow with the level of disruption for all algorithms. For instance, in the disruption level "+10", the performance gaps are never higher than 5\%. But, in the change level "+80" the performance gap are never lower than 11\%. 

In all the evaluated cases, the difference between the Rupture TAR and the Average TAR is higher than the TAR standard deviation. For instance, for the DRL algorithm, in network load disruption level of +50\%, rupture TAR is 17\% lower than Average TAR which is 3.94 times the TAR standard deviation. 

The algorithm with the lower performance gaps is HA-DRL with $\beta=1.0$ as we can see in columns "Rupture TAR - Avg. TAR" and "Rupture TAR - Last TAR" of Table~\ref{tab:ha-drl_1.0_results}. We can state that this algorithm has significantly better robustness then all the others as its performance gaps are significantly lower. However, HA-DRL with $\beta=1.0$ has the worst TAR performance as shown in Fig.~\ref{fig:disruption_levels_1}, \ref{fig:disruption_levels_2} and \ref{fig:disruption_levels_3}, which reduces its applicability.

HA-DRL with $\beta=2.0$ has the second better robustness and  DRL the third as we can see on "Rupture TAR - Avg. TAR" and "Rupture TAR - Last TAR" columns of Tables~\ref{tab:ha-drl_2.0_results} and \ref{tab:drl_results}, respectively. Even if the usage of the Heuristic Function has helped HA-DRL, with $\beta \in \{0.1, 0.5\}$ to achieve significantly better TARs than DRL, the influence of the Heuristic Function in these algorithms was not sufficient to allow to improve the robustness of the DRL algorithm against unpredictable network load disruptions (see Tables~\ref{tab:ha-drl_0.1_results} and \ref{tab:ha-drl_0.5_results}, respectively).


We can observe, however, that HA-DRL with $\beta=2.0$ has better robustness against unpredictable network load changes than DRL as the performance gaps obtained with HA-DRL with $\beta=2.0$ are significantly lower than the ones obtained with DRL as can be observed in columns "Rupture TAR - Avg. TAR" and "Rupture TAR - Last TAR" of Tables~\ref{tab:ha-drl_2.0_results} and \ref{tab:drl_results}, respectively.  These results confirm that HA-DRL with $\beta=2.0$ is the algorithm among those evaluated that is the most adapted to be used in practice. Indeed, the algorithm presents not only the better TAR results and  quick convergence but also robust performance.

\section{Conclusion \label{sec:conclusion}}
We have specifically introduced two DRL-based algorithms and evaluated their performance in a non-stationary network load scenario with unpredictable changes.

In line with the conclusions of 
\cite{HA_DRL_TNSM,cnsm_2021}, the numerical experiments performed in this paper show  that coupling DRL and heuristic functions yields good and stable results even under non stationary load conditions. Therefore, we believe that such an approach is relevant in real networks that are subject to unpredictable network load changes.

As part of our future work, we plan to explore distribution and parallel computing techniques to solve the considered  multi-objective optimization problem using multi-agent or federated learning approaches to address slice placement in  heterogeneous networks mainly when the network is decomposed into several segments or technical domains where the network abstraction introduced in this paper is no more valid. Indeed, each segment should have its own abstractions and data. It is then necessary to share information between the segments to take a global decision. Instead of exchanging complete network states, segments would exchanging minimal information obtained via heuristics. 
 
%

\section*{Acknowledgment}

This work has been performed in the framework of 5GPPP MON-B5G project (www.monb5g.eu). The experiments were conducted using Grid'5000, a large scale testbed by Inria and Sorbonne University (www.grid5000.fr).

%

\bibliographystyle{IEEEtran}
\bibliography{IEEEabrv,my_bib}

\begin{thebibliography}{10}
\providecommand{\url}[1]{#1}
\csname url@samestyle\endcsname
\providecommand{\newblock}{\relax}
\providecommand{\bibinfo}[2]{#2}
\providecommand{\BIBentrySTDinterwordspacing}{\spaceskip=0pt\relax}
\providecommand{\BIBentryALTinterwordstretchfactor}{4}
\providecommand{\BIBentryALTinterwordspacing}{\spaceskip=\fontdimen2\font plus
\BIBentryALTinterwordstretchfactor\fontdimen3\font minus
  \fontdimen4\font\relax}
\providecommand{\BIBforeignlanguage}[2]{{%
\expandafter\ifx\csname l@#1\endcsname\relax
\typeout{** WARNING: IEEEtran.bst: No hyphenation pattern has been}%
\typeout{** loaded for the language `#1'. Using the pattern for}%
\typeout{** the default language instead.}%
\else
\language=\csname l@#1\endcsname
\fi
#2}}
\providecommand{\BIBdecl}{\relax}
\BIBdecl

\bibitem{3GPP}
3GPP, ``{Management and orchestration; 5G Network Resource Model (NRM); Stage 2
  and stage 3 (Release 17)},'' {3rd Generation Partnership Project (3GPP)},
  Technical Specification (TS) 28.541, Dec. 2020, version 17.1.0.

\bibitem{etsi}
\BIBentryALTinterwordspacing
{ETSI NFV ISG}, ``{Network Functions Virtualisation (NFV); Evolution and
  Ecosystem; Report on Network Slicing Support, ETSI Standard GR NFV-EVE 012
  V3.1.1},'' ETSI, Tech. Rep., 2017. [Online]. Available:
  \url{https://www.etsi.org/technologies-clusters/technologies/nfv}
\BIBentrySTDinterwordspacing

\bibitem{survey_vnf_ra_2016}
J.~{Gil Herrera} and J.~F. {Botero}, ``Resource allocation in {NFV}: A
  comprehensive survey,'' \emph{IEEE Trans. Netw. Service Manag.}, vol.~13,
  no.~3, pp. 518--532, Sep. 2016.

\bibitem{survey_vfnp}
A.~Laghrissi and T.~Taleb, ``A survey on the placement of virtual resources and
  virtual network functions,'' \emph{{IEEE} Commun. Surveys Tuts.}, vol.~21,
  no.~2, pp. 1409--1434, 2nd. Quart., 2019.

\bibitem{netsoft_2020}
J.~J.~A. Esteves, A.~Boubendir, F.~Guillemin, and P.~Sens, ``Location-based
  data model for optimized network slice placement,'' in \emph{Proc. 2020 6th
  IEEE Conf. Netw. Softwarization (NetSoft)}, 2020, pp. 404--412.

\bibitem{vne_np_hardness}
E.~Amaldi, S.~Coniglio, A.~M. Koster, and M.~Tieves, ``On the computational
  complexity of the virtual network embedding problem,'' \emph{Electron. Notes
  Discrete Math.}, vol.~52, pp. 213--220, Jun. 2016.

\bibitem{p1}
Z.~{Yan}, J.~{Ge}, Y.~{Wu}, L.~{Li}, and T.~{Li}, ``Automatic virtual network
  embedding: A deep reinforcement learning approach with graph convolutional
  networks,'' \emph{{IEEE} J. Sel. Areas Commun.}, vol.~38, no.~6, pp.
  1040--1057, Jun. 2020.

\bibitem{p2}
M.~{Dolati}, S.~B. {Hassanpour}, M.~{Ghaderi}, and A.~{Khonsari}, ``{DeepViNE}:
  Virtual network embedding with deep reinforcement learning,'' in \emph{Proc.
  IEEE INFOCOM 2019 - IEEE Conf. Comput. Commun. Workshops (INFOCOM WKSHPS)},
  2019, pp. 879--885.

\bibitem{p5}
H.~Yao, X.~Chen, M.~Li, P.~Zhang, and L.~Wang, ``A novel reinforcement learning
  algorithm for virtual network embedding,'' \emph{Neurocomputing}, vol. 284,
  pp. 1--9, Apr. 2018.

\bibitem{p3}
H.~Wang, Y.~Wu, G.~Min, J.~Xu, and P.~Tang, ``Data-driven dynamic resource
  scheduling for network slicing: A deep reinforcement learning approach,''
  \emph{Inf. Sci.}, vol. 498, pp. 106--116, Sep. 2019.

\bibitem{p4}
Y.~{Xiao}, Q.~{Zhang}, F.~{Liu}, J.~{Wang}, M.~{Zhao}, Z.~{Zhang}, and
  J.~{Zhang}, ``{NFVdeep}: Adaptive online service function chain deployment
  with deep reinforcement learning,'' in \emph{Proc. 2019 IEEE/ACM 27th Int.
  Symp. Qual. Service (IWQoS)}, 2019, pp. 1--10.

\bibitem{p8}
P.~T.~A. {Quang}, A.~{Bradai}, K.~D. {Singh}, and Y.~{Hadjadj-Aoul},
  ``Multi-domain non-cooperative {VNF-FG} embedding: A deep reinforcement
  learning approach,'' in \emph{Proc. IEEE INFOCOM 2019 - IEEE Conf. Comput.
  Commun. Workshops (INFOCOM WKSHPS)}, 2019, pp. 886--891.

\bibitem{sutton2018reinforcement}
R.~S. Sutton and A.~G. Barto, \emph{Reinforcement learning: An
  introduction}.\hskip 1em plus 0.5em minus 0.4em\relax Cambridge, MA, USA: MIT
  press, 2015.

\bibitem{HA_DRL_TNSM}
J.~J. {Alves Esteves}, A.~{Boubendir}, F.~{Guillemin}, and P.~{Sens}, ``A
  heuristically assisted deep reinforcement learning approach for network slice
  placement,'' \emph{arXiv preprint arXiv:2105.06741}, 2021.

\bibitem{cnsm_2021}
J.~J.~A. Esteves, A.~Boubendir, F.~Guillemin, and P.~Sens, ``Drl-based slice
  placement under non-stationary conditions,'' \emph{Submitted to IEEE 17th
  International Conference on Network and Service Management (CNSM)}, 2021.

\bibitem{quang2019deep}
P.~T.~A. Quang, Y.~Hadjadj-Aoul, and A.~Outtagarts, ``A deep reinforcement
  learning approach for vnf forwarding graph embedding,'' \emph{IEEE Trans.
  Netw. Service Manag.}, vol.~16, no.~4, pp. 1318--1331, Dec. 2019.

\bibitem{rkhami2021learn}
A.~{Rkhami}, Y.~{Hadjadj-Aoul}, and A.~{Outtagarts}, ``Learn to improve: A
  novel deep reinforcement learning approach for beyond {5G} network slicing,''
  in \emph{Proc. 2021 IEEE 18th Annu. Consum. Commun. Netw. Conf. (CCNC)},
  2021, pp. 1--6.

\bibitem{new_1}
J.~Pei, P.~Hong, M.~Pan, J.~Liu, and J.~Zhou, ``Optimal vnf placement via deep
  reinforcement learning in sdn/nfv-enabled networks,'' \emph{{IEEE} J. Sel.
  Areas Commun.}, vol.~38, no.~2, pp. 263--278, Feb. 2020.

\bibitem{bertsimas2006robust}
D.~Bertsimas and A.~Thiele, ``Robust and data-driven optimization: modern
  decision making under uncertainty,'' in \emph{Models, methods, and
  applications for innovative decision making}.\hskip 1em plus 0.5em minus
  0.4em\relax INFORMS, 2006, pp. 95--122.

\bibitem{shafique2020robust}
M.~Shafique, M.~Naseer, T.~Theocharides, C.~Kyrkou, O.~Mutlu, L.~Orosa, and
  J.~Choi, ``Robust machine learning systems: Challenges, current trends,
  perspectives, and the road ahead,'' \emph{IEEE Design \& Test}, vol.~37,
  no.~2, pp. 30--57, Apr. 2020.

\bibitem{al2021robustness}
R.~R.~O. Al-Nima, T.~Han, S.~A.~M. Al-Sumaidaee, T.~Chen, and W.~L. Woo,
  ``Robustness and performance of deep reinforcement learning,'' \emph{Applied
  Soft Comput.}, vol. 105, p. 107295, Jul. 2021.

\bibitem{marotta2017fast}
A.~Marotta, E.~Zola, F.~D'Andreagiovanni, and A.~Kassler, ``A fast robust
  optimization-based heuristic for the deployment of green virtual network
  functions,'' \emph{J. Netw. Comput. Applications}, vol.~95, pp. 42--53, Jul.
  2017.

\bibitem{marotta2017energy}
A.~Marotta, F.~D’andreagiovanni, A.~Kassler, and E.~Zola, ``On the energy
  cost of robustness for green virtual network function placement in 5g
  virtualized infrastructures,'' \emph{Comput. Netw.}, vol. 125, pp. 64--75,
  Apr. 2017.

\bibitem{reddy2016robust}
V.~S. Reddy, A.~Baumgartner, and T.~Bauschert, ``Robust embedding of
  vnf/service chains with delay bounds,'' in \emph{2016 IEEE Conf. Netw. Funct.
  Virtualization Softw. Defined Netw. (NFV-SDN)}.\hskip 1em plus 0.5em minus
  0.4em\relax IEEE, 2016, pp. 93--99.

\bibitem{baumgartner2017}
A.~Baumgartner, T.~Bauschert, A.~A. Blzarour, and V.~S. Reddy, ``Network slice
  embedding under traffic uncertainties — a light robust approach,'' in
  \emph{2017 13th Int. Conf on Netw. Service Manag. (CNSM)}.\hskip 1em plus
  0.5em minus 0.4em\relax IEEE, 2017, pp. 1--5.

\bibitem{robust_1}
P.~Sun, J.~Lan, J.~Li, Z.~Guo, and Y.~Hu, ``Combining deep reinforcement
  learning with graph neural networks for optimal vnf placement,'' \emph{IEEE
  Commun. Letters}, vol.~25, no.~1, pp. 176--180, Jan. 2021.

\bibitem{slim2018close}
F.~{Slim}, F.~{Guillemin}, and Y.~{Hadjadj-Aoul}, ``{CLOSE}: A costless service
  offloading strategy for distributed edge cloud,'' in \emph{Proc. 2018 15th
  IEEE Annu. Cons. Commun. Netw. Conf. (CCNC)}, 2018, pp. 1--6.

\bibitem{farah2}
F.~{Slim}, F.~{Guillemin}, A.~{Gravey}, and Y.~{Hadjadj-Aoul}, ``Towards a
  dynamic adaptive placement of virtual network functions under {ONAP},'' in
  \emph{Proc. 2017 IEEE Conf. on Netw. Function Virtualization Softw. Defined
  Netw. (NFV-SDN)}, 2017, pp. 210--215.

\bibitem{a3c}
V.~Mnih, A.~P. Badia, M.~Mirza, A.~Graves, T.~Lillicrap, T.~Harley, D.~Silver,
  and K.~Kavukcuoglu, ``Asynchronous methods for deep reinforcement learning,''
  in \emph{Int. Conf. Mach. Learn.}\hskip 1em plus 0.5em minus 0.4em\relax
  PMLR, 2016, pp. 1928--1937.

\bibitem{kipf_gcn}
T.~N. Kipf and M.~Welling, ``Semi-supervised classification with graph
  convolutional networks,'' in \emph{Proc. 5th Int. Conf. Learn.
  Representations (ICLR)}, 2017, pp. 1--14.

\bibitem{cnsm_2020}
J.~J. {Alves Esteves}, A.~{Boubendir}, F.~{Guillemin}, and P.~{Sens},
  ``Heuristic for edge-enabled network slicing optimization using the “power
  of two choices”,'' in \emph{Proc. 2020 IEEE 16th Int. Conf. Netw. Service
  Manag. (CNSM)}, 2020, pp. 1--9.

\end{thebibliography}

%
%

\begin{IEEEbiography}[{\includegraphics[width=1in,height=1.25in,clip,keepaspectratio]{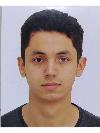}}]{José~Jurandir~Alves~Esteves} graduated from the University of Clermont Auvergne in 2017 and from the Federal University of Minas Gerais in 2019 obtaining two engineering degrees and a master's degree in computer science from the University of Clermont Auvergne. He is doing his PhD between Orange Labs and the Computer Science Laboratory of Paris 6 (LIP6), Sorbonne University. His research is related to automation and optimization models and algorithms for network slice orchestration.
\end{IEEEbiography}

\begin{IEEEbiography}[{\includegraphics[width=1in,height=1.5in,clip,keepaspectratio]{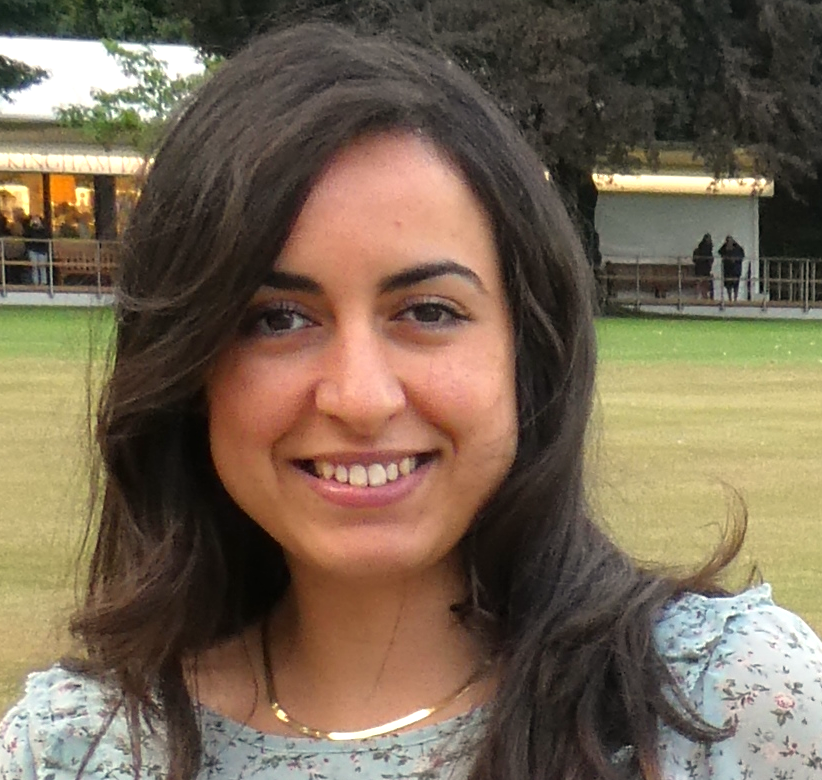}}]
{Amina Boubendir} is a researcher and project manager at Orange Labs in France. Her research is in the area of design, management and softwarization of networks and services. Amina received a Master degree in Network Design and Architecture from Télécom Paris in 2013, and a PhD in Networking and Computer Science from Télécom Paris in 2016. She is a member of the Orange Expert community on "Networks of the Future".
\end{IEEEbiography}

\begin{IEEEbiography}[{\includegraphics[width=1in,height=1.25in,clip,keepaspectratio]{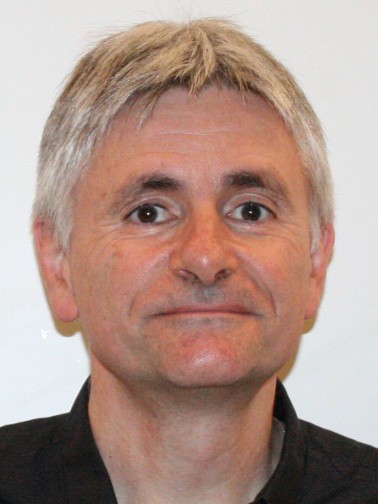}}]
{Fabrice Guillemin}
graduated from Ecole Polytechnique in 1984 and from Telecom Paris in 1989. He received the PhD degree from the University of Rennes in 1992. He defended his “habilitation” thesis in 1999 at the University Pierre et Marie Curie (LIP6), Paris. Since 1989, he has been with Orange Labs (former CNET and France Telecom R\&D). He is currently leading a project on the evolution of network control. He is a member of the Orange Expert community on "Networks of the Future".
\end{IEEEbiography}


\begin{IEEEbiography}[{\includegraphics[width=1in,height=1.25in,clip,keepaspectratio]{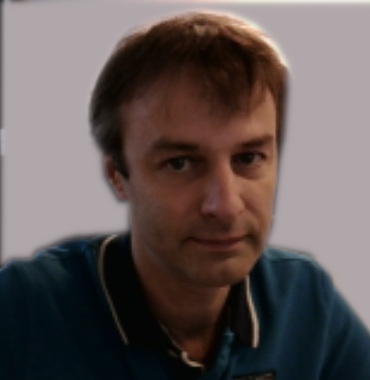}}]{Pierre Sens}
received his Ph.D. in Computer Science in 1994, and the “Habilitation à diriger des recherches” in 2000 from Paris 6 University (UPMC), France. Currently, he is a full Professor at Sorbonne Université (ex-UPMC). His research interests include distributed systems and algorithms, large scale data storage, fault tolerance, and cloud computing. He is leading Delys a joint research team between LIP6 and Inria. He was member of the Program Committee of major conferences in the areas of distributed systems and parallelism (ICDCS, IPDPS, OPODIS, ICPP, Europar, SRDS, DISC. . . ) and has served as general chair of SBAC-PAD and EDCC. Overall, he has published over 150 papers in international journals and conferences and has acted for advisor of 25 Ph.D. thesis.
\end{IEEEbiography}

\flushend

\end{document}